\documentstyle[12pt]{article}

\begin{document}

\title{\bf Evidence for Nonvanishing Cosmological Constant\\
in NonSusy Superstring Models}

\author{Roberto Iengo\thanks{E-mail: iengo@he.sissa.it}\\
{\it Scuola Internazionale Superiore di Studi Avanzati (SISSA)} \\
{\it and INFN, Sezione di Trieste}\\
{\it Via Beirut 4, I-34013 Trieste, Italy}
\\
\\
Chuan-Jie Zhu\thanks{E-mail: zhucj@itp.ac.cn} \\
{\it Institute of Theoretical Physics, Chinese Academy of}\\
{\it Sciences, P. O. Box 2735, Beijing 100080, P. R. China}}

\maketitle

\begin{abstract}
We reanalyse the computation of the  cosmological constant $\Lambda$ at two
loops in recently proposed Superstring models without massless gravitini,
both in the theta-function based formalism and by a detailed computation
in the more explicit hyperelliptic description of the underlying genus two
Riemann surface. 
$\Lambda$ is expressed as the integral over the surface moduli of
an amplitude which is zero if susy is not completely broken, but we find
it to be nonvanishing in the susy breaking models which can be given an
explicitly workable fermionic formulation. Thus unfortunately the issue
of getting realistic and perturbatively viable models from Superstring    
Theory remains open.

\end{abstract}

\newpage

\section{Introduction and summary}

The vanishing of the cosmological constant $\Lambda$ is a crucial issue in
Superstring
theory. In fact, if  $\Lambda$  is nonzero the theory
is perturbatively unstable. Unavoidable (infrared) divergences
appear due to the dilaton tadpole (proportional to $\Lambda$), 
physically signaling the fact that the flat background, over
which the perturbative expansion is performed, is no longer consistent.
Therefore also the perturbative spectrum and thus the particle content of
the theory is no longer trusty, and its physical meaning becomes 
(at least so far) inaccessible. 

The standard versions of Superstring theory have indeed zero cosmological
constant \cite{GSW,Joe}. This can be formally argued as a consequence of exact
supersymmetry, and it has been explicitly proven at one and two 
loops \cite{GavaIengoSotkov},
both for the maximally supersymmetric models and also for less
(but still at least $N=1$) supersymmetry \cite{IengoGalenI}.
It is therefore very important to investigate whether it is possible to
construct more realistic string theories, where 
there are no massless gravitini and supersymmetry is 
completely broken. Models of that kind have recently appeared in the
literature \cite{KachruSilversteinA,TyeA}, 
and including arguments to show that $\Lambda$  is
still zero \cite{KachruSilversteinB}.
Some of those models can be given an  explicitly workable
fermionic formulation \cite{TyeA}, by using which it is possible to make a
detailed analysis also at two loops. 
Note that the one and two loop computations are physically
very different.
In fact at one loop one evaluates the partition function of a free theory,
whereas only at two loops one begins to see the effect of the interaction.

Here we reanalyse those models.
Our rather explicit computation will make use of the very
convenient
hyperelliptic formalism which provides a complete description of the genus
two Riemann surface. Thus it looks different from the formalism used in the
argument appeared in the recent literature about the two loops vanishing
of the $\Lambda$  in the nonsusy models. However
we have also reanalysed those arguments
\cite{TyeA,KachruSilversteinB} and found some flaws.
One of the main ingredients of the two loop computation 
in any formalism is the 
occurrence in the amplitude of a correlator of supercurrents,
related to the fact that at genus two there are superghost zero modes also
for the even spin structures. Now, in some arguments this correlator is
ignored \cite{TyeA}
(note that the correlator is sigular at coincident points);  
if one does so the result is indeed zero (as we check in
our explicit computation, see eq. (\ref{diffbb})). 
However the amplitude is not the right one.

In more refined arguments this correlator is included. Let us see it more
precisely  (see for instance \cite{VerlindeTrieste}). 
The amplitude $A$, to be integrated over the moduli of the surface, 
is obtained by evaluating the
expectation value of the following quantity (times determinants of some
differential operators): 
\begin{equation}
\delta (\beta (z_1))\psi^{\mu}\partial X_{\mu}(z_1)
\delta (\beta (z_2)) \psi^{\mu}\partial X_{\mu}(z_2).
\end{equation}
This is eq. (2.10) of \cite{VerlindeTrieste}, taking 
the supercurrent insertion in a point: $\chi_a =\delta (z_a)$,
and considering the matter part $J_{m}=\psi^{\mu}\partial X_{\mu}$
of the supercurrent. In the full expression one has
to complete the supercurrent by adding the ghost part and the
compactified matter part  to the matter part, 
but for the purpose of the short discussion
here it is enough to look at the partial expression above.

The result is (up to theta-function factors):
\begin{equation}
A= (\hbox{theta-functions})\frac
{k_{ij}\omega_i (z_1)\omega_j
(z_2)+q\partial_{z_1}\partial_{z_2}\log E(z_1,z_2)}
{E(z_1,z_2)^2 \sigma (z_1)^2\sigma (z_2)^2},
\label{eqtwo}
\end{equation}
where $k_{ij}$ and $q$ are constants. This is obtained by
using the bosonization eqs. (3.15) and (3.30) of \cite{VerlindeTrieste}:
\begin{equation}
\langle \delta (\beta (z_1))\delta (\beta (z_2))\rangle=(\hbox{theta-functions})
/E(z_1,z_2)\sigma (z_1)^2\sigma (z_2)^2,
\end{equation}
and by computing 
\begin{equation}
\langle J_{m}(z_1)J_{m}(z_2)\rangle =(\hbox{theta-functions})\, 
{ (k_{ij}\omega_i (z_1)\omega_i
(z_2)+q\partial_{z_1}\partial_{z_2}\log E(z_1,z_2))\over E(z_1,z_2)} .
\end{equation}

Note that the amplitude $A$ has to be a 0-differential in $z_{1,2}$,
in order to be independent (up to irrelevant total derivatives in
the moduli) on those arbitrary points. 
Remembering that $\sigma$ is a differential of degree equal to half the genus
thus $\sigma^2$ is a 2 differential in our case (genus 2), 
that $E$ is a $-1/2$
differential
and that $\omega$ like $\partial \log E$ is a $1$-differential, we see that the
result is in fact a $0$-differential in $z_{1,2}$, indeed the same $0$-degree
as  $\delta (\beta )\psi^{\mu}\partial X_{\mu}$.

Now, in eq. (5.1) of ref. \cite{KachruSilversteinB} (the key point of that
argument)  
the formula for $A$, eq. (\ref{eqtwo}), appears
with a  crucial modification: that is, there is the extra factor 
$\det(\Phi^{3/2}_a (z_b))$, where $\Phi^{3/2}_a$ are the superghost zero mode
functions which are $3/2$-differentials. 
(Instead, the factor $1/\sigma (z_1)^2\sigma (z_2)^2$ has not
been written there, 
but since it is spin structure independent and nonzero for
$z_1\to z_2$ it is not relevant for the ensuing argument).
But actually there is not such a factor $\det(\Phi^{3/2}_a (z_b) )$. Indeed,
as a check, since it is a half-integer differential,  it does not match
with the previously recalled property of $A$ to be a $0$-differential.
If we remove this factor to get the right amplitude, 
we see that the argument of ref. \cite{KachruSilversteinB} does no longer 
show the vanishing of $\Lambda$, apart from possibly
the ``simpler model" of ref. \cite{KachruSilversteinB} 
for which however the explicit form of the
partition function and the explicit results for the spectrum
have not been given, and the fundamental integration region over the moduli
appears to be nonstandard.

This is confirmed by our explicit two loop  computation for those
models which can be given an explicitly workable free fermion formulation. 
First, in Sect. 2, we review the fermionic construction and the model 
(whose gravitini content is explicitly analysed in Appendix A) and
in Sect. 3, in some details, the one loop computation
showing that $\Lambda$ is zero at this order. 
The one loop computation  indicates a convenient strategy for the two loop  one.
Accordingly, in Sect.4, the two loop 
amplitude is conveniently organized as a sum of various terms,
which in Appendix B are classified as modular invariant sets.
Many of those terms, Sect.4.1, are the same as for supersymmetric models
and they were shown to vanish explicitly in ref. \cite{IengoGalenI} (see also
\cite{ABKb}),
(we review this computation 
in the Appendix C following the strategy of one loop computation).
Besides them there remain two sets of terms,
which only occur in nonsusy models. 
Those of the first set, Sect.4.2,  are shown to vanish by essentially the same
kind
of computation as at one loop (that is, showing that they contain at least
one odd theta function). We finally, Sect.4.3, focus on the remaining set:
here we find a nonvanishing result (the crucial expression is 
eq. (\ref{diffbb}),
where we see that contrary to the susy case there is a crucial change of sign
which prevent the cancellation). We complete the discussion by
making the complete sum over the spin structures, to get a modular
invariant result which contributes
to the integrand of two-loop cosmological cosntant,
eq. (\ref{ModularInvariant}). 

Thus unfortunately we find that at two loops
the same computation showing that the cosmological constant is zero
when supersymmetry is unbroken gives a result different from zero
in absence of supersymmetry, at least for those models which we have been
able to work out explicitely. To be precise, this computation is done
for the amplitude before the integration over the moduli of the genus two 
Riemann surface. However all the proofs appeared so far have been done
in this way (including those under question for the nonsusy string theories),
and indeed there is no indication that the final integration over the moduli
should vanish. Thus our conclusion is that the issue of getting
realistic and viable models from Superstring is still an unsolved,
possibly the main unsolved problem in this context.

\section{A review of the free fermionic construction of type 
II superstring  theories}

In this section we review the free fermionic construction of lower dimensional
superstring models \cite{ABK,Tye}. 
The presentation is tailored to the type II superstring theory.

For type II superstring theory in $D$ dimensions, we have the
following list of two-dimensional fields appearing in the
construction of the models:

$X^{\mu},~\psi^{\mu}, ~\tilde{\psi}^{\mu}$ 
($\mu =0,1,\cdots,D-1$) which have a
space-time interpretaion;

$\chi^I,y^I,\omega^I$, $I =1, \cdots, 10-D$ which are
$3(10-D)$ free left (holomorphic) Majorana
fermions; 

$ \tilde{\chi}^I,\tilde{y}^I,\tilde{\omega}^I$, 
$I =1, \cdots, 10-D$ which are
$3(10-D)$ free right  (anti-holomor-phic) Majorana
fermions.

\noindent The left  supercurrent is realized nonlinearly by
these fields as \cite{ABKW}
\begin{eqnarray}
J & = & \psi\cdot \partial X + 2 \, c \, \partial \beta - \gamma \, 
b + 3 \, \partial c\, \beta +  \sum_{I=1}^{10-D}   
\chi^I\,y^I\,\omega^I, 
\nonumber \\
&\equiv & J_m + J_{{gh}} + J_{\chi} , 
 \label{eqone}
\end{eqnarray}
and similarly for the right  supercurrent $\tilde{J}$.
As it is well-known, fermionic fields have non-trivial boundary
conditions. For different boundary conditions these fermionic
fields give different contributions to the partition function.
The partition function should be modular invariant, guarantee
the cancellation of gauge and gravitational anomalies (in the
low energy or field theory limit of string theory). How modular
invariance restricts the possible choice of boundary conditions
was discussed in \cite{ABK,Tye}. By studying the one-loop
partition function and its multi-loop counterparts, they found
a set of rules of constructing modular invariant string theories.
By using their rules on can construct a lot of consistent models.
Let's  briefly review their construction, setting $D=4$ in
what follows.

To start with let us first recall some notations. For the torus it
can be represented by a flat parallelogram in the complex plane
with side 1 and $\tau$ corresponding to its two non-contractible
loops. We denote the one-loop spin structure as 
$\Big[  \begin{array}{c}  \alpha \cr
\beta  \end{array} \Big]$
where $\alpha$ and $\beta$ are subsets of 
$F$ ($\equiv$ the set of all the fermions) containing those
fermions that are periodic around 1 and $\tau$ respectively. For
any set $\alpha$ of fermions ($\alpha \subseteq F$)
we define its characteric function as
\begin{equation}
\alpha(f) = 1, \qquad \hbox{if} \quad f\in \alpha; 
\qquad  \alpha(f)=0, \qquad \hbox{otherwise}. 
\end{equation}
We also define addition and multiplication of fermion sets as
ordinary addition and multiplication modulo 2 of their
characteristic functions. Thus addition is the symmetric difference
and multiplication the intersection:
\begin{eqnarray}
& & \alpha + \alpha'  = \alpha \cup \alpha' - \alpha \cap \alpha', 
\\
& & \alpha \alpha' =  \alpha \cap \alpha' .
\end{eqnarray}
The one-loop vacuum amplitude can be written as
\begin{equation}
Z_{one-loop} = \int { d^2 \tau \over (\hbox{Im} \tau)^{ 
{D\over 2 } + 1 } |\eta|^{24} }
\sum_{\alpha , \beta} C\Big[  \begin{array}{c}
\alpha \cr \beta
\end{array} \Big]\, 
\prod_{f \in F'} \Theta^{1/2}  
\Big[\begin{array}{c}
\alpha(f)\cr \beta(f)
\end{array} \Big](\tau). \label{oneloopa}
\end{equation}
Here $\eta(\tau)$ is the Didekin eta function; 
$\Theta \Big[ \begin{array}{c}
\epsilon \cr \epsilon'  \end{array} \Big] $
is the Jacobi theta function with characteristic 
$\Big[ \begin{array}{c}
\epsilon/2 \cr \epsilon' /2 \end{array} \Big] $.
By slight abuse of notation we suppress the fact that this theta
function should be complex conjugated when $f$ is a right
(anti-holomorphic) fermion. For later convenience, we also define 
$F'$ as  the set of all ``transverse''
fermions, i.e. $F$ minus two $\psi^{\mu}$'s and
two $\tilde{\psi}^{\mu}$'s. ( For later convenience, we also
 use $F_{\mathrm{L}}$
and $F_{\mathrm{R}}$ to denote all the left fermions and right
fermions respectively. $F_{\mathrm{L}}'$ to denote all the left
fermions minus two $\psi^{\mu}$'s and similarly for 
$F_{\mathrm{R}}'$). 

It is important to note that in the 1-loop case there are no 
superghost zero modes (in the even spin structures; the odd spin structure 
gives trivially a vanishing contribution to the cosmological constant).
Thus, in the expression for $Z_{one-loop}$ there is no factor 
containing the supercurrent (or picture changing operators) correlator.
Instead, in the two loops case, since there are superghost zero modes also
for the even spin structures, a supercurrent correlator will appear 
inside the sum over the spin structures defining $Z_{two-loops}$.
This correlator will play a crucial role in the issue whether the
cosmological constant is also zero at two loops.  

Finally the coefficients $C\Big[ \begin{array}{c} \alpha \cr
\beta \end{array} \Big]$ takes values equal to either $+1$ 
or $-1$ which should be determined from modular invariance. 

By modular invariance, the following conditions on the
coefficients  $ C\Big[ \begin{array}{c} 
\alpha \cr \beta  \end{array} \Big] $ 
are derived \cite{ABK,Tye}:
{\arraycolsep=3pt
\begin{eqnarray}
& & C\Big[ \begin{array}{c}
\alpha \cr \beta  \end{array} \Big]  =  \varepsilon_{\alpha} 
\, C\Big[ \begin{array}{c}
\alpha \cr \alpha + \beta + F   \end{array} \Big], 
\label{conditiona}
\\
& & C\Big[ \begin{array}{c}
\alpha \cr \beta  \end{array} \Big]  = 
 \varepsilon_{\alpha\beta }^2
\, C\Big[ \begin{array}{c}
\beta \cr \alpha \end{array} \Big], 
\\
& & C\Big[ \begin{array}{ccc}
\alpha_1 & \cdots  & \alpha_g  \cr \beta_1 & 
\cdots & \beta_g   \end{array} \Big]  =  
C\Big[ \begin{array}{c}
\alpha_1  \cr \beta_1  \end{array} \Big] \cdots 
C\Big[ \begin{array}{c}
\alpha_g  \cr \beta_g  \end{array} \Big] , 
\\
& & C\Big[ \begin{array}{c}
\alpha \cr  \beta  \end{array} \Big]
C\Big[ \begin{array}{c}
\alpha'\cr  \beta'  \end{array} \Big] 
= (-1)^{\alpha(\psi,\tilde{\psi}) +\alpha'(\psi,\tilde{\psi})  } \,
\varepsilon^2_{\alpha\alpha'} \,
C\Big[ \begin{array}{c}
\alpha \cr  \beta  + \alpha' \end{array} \Big]
C\Big[ \begin{array}{c}
\alpha'\cr  \beta'   + \alpha \end{array} \Big], 
\label{conditiond}
\end{eqnarray}
where 
$\alpha(\psi,\tilde{\psi}) = \alpha(\psi) + \alpha(\tilde{\psi})$,
$\varepsilon_{\alpha} = \exp( { i \pi n(\alpha) \over 8} )$,
etc. and $n(\alpha)$ is the difference between the number of
left fermions 
in the set $\alpha$ (not counting the two longitudinal fermions if 
$\psi^{\mu}$'s and/or $\tilde{\psi}^{\mu}$'s are in $\alpha$, see below)
and the number  of right fermions in $\alpha$.
These equations are all the equations required for modular
invariance to all loops. }

The analysis of these equations is rather lengthly \cite{ABK},
but the results can be presented with reasonable dispatch. The sets
of fermions which enter the summation in eq. (\ref{oneloopa}) 
form an additive 
group $\Xi$ of subsets of $F$, containing in particular $F$ and the 
empty set $\emptyset$. The precise choice of $\Xi$ and 
$C\Big[ \begin{array}{c} \alpha \cr  \beta  \end{array} \Big]$'s
define the theory. This choice 
is restricted by a series of consistency conditions, which are
derived from eqs. (\ref{conditiona})--(\ref{conditiond}). 
Firstly, admissible spin structure assignments to the fermions must be
such that the spercurrent $J$ (and also for $\tilde{J}$) has also a
well-defined spin structure. In particular this implies that for any 
assignment $C\Big[ \begin{array}{c} \alpha \cr  \beta  \end{array} \Big]$
the supercurrent $J$ and the superghosts, as well as the $D$
fermion field $\psi^{\mu}$'s all have the same spin structure
$\Big[ \begin{array}{c} \alpha(\psi^{\mu}) \cr
\beta(\psi^{\mu})   \end{array} \Big]$. All the $D$ fermion field
$\psi^{\mu}$'s are either all in the set $\alpha$ or all not in the 
set.  The second condition is that for
all $\alpha, \alpha'$, etc. $\in \Xi$:
\begin{eqnarray}
& & n(\alpha) = 0~\hbox{mod} 8,
\nonumber \\
& & n(\alpha\alpha') = 0~\hbox{mod} 4,
\label{seta}
\\
& &  n(\alpha\alpha'\alpha''\alpha''') = 0~\hbox{mod} 2.
\nonumber 
\end{eqnarray}
Finally the coefficients 
$C\Big[ \begin{array}{c} \alpha \cr  \beta  \end{array} \Big]$
must obey
\begin{eqnarray}
C\Big[ \begin{array}{c} \alpha \cr  \beta  \end{array} \Big]
& = & \epsilon_{\alpha\beta}^2 
C\Big[ \begin{array}{c} \beta  \cr  \alpha   \end{array} \Big],
\nonumber \\
C\Big[ \begin{array}{c} \alpha \cr  \alpha  \end{array} \Big]
& = & \epsilon_{\alpha}
C\Big[ \begin{array}{c} \alpha  \cr  F   \end{array} \Big],
\label{setb}
 \\
C\Big[ \begin{array}{c} \alpha \cr  \beta + \gamma \end{array} \Big] 
& = & (-1)^{\alpha(\psi,\tilde{\psi})}
\, C\Big[ \begin{array}{c} \alpha \cr  \beta  \end{array} \Big]
\, C\Big[ \begin{array}{c} \alpha \cr  \gamma  \end{array} \Big].
\nonumber 
\end{eqnarray}
Note that conditions (\ref{seta}) ensure the consistency of
eq. (\ref{setb}).  For our later use one can derive the following 
useful formula:
\begin{equation}
C\Big[ \begin{array}{c}
\alpha + \beta  \cr  \gamma  \end{array} \Big] = (-1)^{ \gamma(\psi^{\mu},
\tilde{\psi}^{\mu}) } \,\exp\Big\{ {i\pi \over 2 }\, n(\alpha \beta  \gamma) \Big\} \, 
C\Big[ \begin{array}{c}
\alpha \cr  \gamma  \end{array} \Big] 
C\Big[ \begin{array}{c}
 \beta  \cr  \gamma  \end{array} \Big] . 
\end{equation}

By carefully analysing the spectrum of non-interacting string 
states,  it was found that a string model has a massless gravitino
 if and   only  if there exists a set $S\in \Xi$ such that the
following two  conditions  are satisfied \cite{ABK}:

1. $S$ is a set of precisely eight left fermions including 
$\psi^{\mu}$ (not counting the two longitudinal fermions)
or a set of precisely eight right fermions including 
$\tilde{\psi}^{\mu}$;

2. for any set $X\in \Xi$ satisfying $X\cap S = \emptyset$, 
we must   have
\begin{equation}
C\Big[ \begin{array}{c}
S \cr  X  \end{array} \Big] = -(-1)^{X(\psi^{\mu}, 
\tilde{\psi}^{\mu}) }. 
\end{equation}

 The model we will study in detail in this paper is constructed in
\cite{TyeA} and it is given by the following generating sets:
\begin{eqnarray}
b_0 = F & = & \{\psi^{\mu=0,\cdots,3}, \chi^{I=1,\cdots,6},
y^{I=1,\cdots,6}, 
\omega^{I=1,\cdots,6}, 
\nonumber \\
& & \qquad \tilde{\psi}^{\mu=0,\cdots,3}, \tilde{\chi}^{I=1,
\cdots,6},
\tilde{y}^{I=1,\cdots,6}, \tilde{\omega}^{I=1,\cdots,6}\},
\nonumber \\
b_1 & = & \{\psi^{\mu=0,\cdots,3}, \chi^{I=1,\cdots,6} \},
\nonumber \\
b_2 & = & \{\tilde{\psi}^{\mu=0,\cdots,3}, \tilde{\chi}^{I=1,
\cdots,6} \}, 
\label{eq18} \\
b_3 & = & \{ \chi^{I=1,\cdots,4}, y^{I=1,\cdots,4}, 
\tilde{y}^{I
=1,\cdots,4}, \tilde{\omega}^{I=1,\cdots,4} \},
\nonumber \\
b_4 & = & \{  {y}^{I
=1,\cdots,4}, \omega^{I=1,\cdots,4}, 
\tilde{\chi}^{I=1,\cdots,4}, \tilde{y}^{I=1,\cdots,4}\}.
\nonumber 
\end{eqnarray}
To completely fix the model we must specify what all these
 coefficients 
$C\Big[ \begin{array}{c}
\alpha \cr  \beta  \end{array} \Big] $'s are. The strategy is the
 following: 
all models obtained by omitting one or two generating sets 
$b_3$ and/or
$b_4$ are supersymmetric. In particular the model generated
 by the sets
 $b_{0,1,2}$ has $(4,4)$ supersymmetry. This requires:
\begin{equation}
C\Big[ \begin{array}{c}
b_1  \cr  b_2 \end{array} \Big] =  C\Big[ \begin{array}{c}
b_2  \cr  b_1 \end{array} \Big] = 1. 
\end{equation}
(See Appendix A.) Notice also that 
\begin{equation}
b_1 \cap b_4 = \emptyset, \qquad \hbox{and} \qquad 
b_2 \cap b_3 = \emptyset,
\end{equation}
from  the general analysis, we will have a
non-supersymmetric model if we have
\begin{equation}
C\Big[  \begin{array}{c} b_1 \cr
b_4 \end{array} \Big] = 1, \qquad \hbox{and} \qquad
C\Big[  \begin{array}{c}
b_2 \cr b_3 \end{array} \Big] = 1.
\label{coefficient}
\end{equation}
We will not analyse the spectrum of this model \cite{TyeA} here. 
An analysis of the gravitini is given in Appendix A. 
For this model it was proved that the one-loop cosmological 
constant is 
zero even before the integration over the moduli space
 parameter $\tau$. 
We will give a complete proof in the next section. The
 method will be used 
in our explicit computation of the two-loop cosmological
constant (which  is non-vanishing before the integration over the moduli space).

\section{One-loop cosmological constant}
 
The one-loop cosmological constant was given in (\ref{oneloopa}).
What we will prove is that the cosmological constant is zero 
even before the integration over the moduli $\tau$. So we can
first consider the following expression:
\begin{eqnarray}
V  & = & \sum_{\alpha, \beta \in \Xi} 
C\Big[  \begin{array}{c} \alpha  \cr
\beta  \end{array} \Big]\, \prod_{f\in F'} \Theta^{1/2}  
\Big[  \begin{array}{c} \alpha(f) \cr
\beta(f) \end{array} \Big](\tau) 
\nonumber \\
& = & \sum_{\alpha, \beta \in \Xi_0}  \sum_{n, m, \tilde{n}, 
\tilde{m}=0 }^1
C\Big[  \begin{array}{c} \alpha + n b_3 + {m}b_4  \cr
\beta  + \tilde{n} b_3 + \tilde{m}b_4  \end{array} \Big]
\nonumber \\
& & \quad \times  \prod_{f\in F'} \Theta^{1/2} 
\Big[  \begin{array}{c} (\alpha+ n b_3 + {m}b_4  )(f) \cr
(\beta + \tilde{n} b_3 + \tilde{m}b_4 ) (f) \end{array} 
\Big]^{1/2} (\tau) .
\end{eqnarray}
In the above expression  we have separated the summation 
over $b_3$ and 
$b_4$ from the summation over $\Xi_0$ which is defined as
 the set generated 
by the sets $b_{0,1,2}$: $\Xi_0 = \{ \sum_{i=0}^2 n_i b_i | 
n_i=0,1\}$. 
Now setting $\alpha \to 
\alpha + n_1 b_1 + m_1 b_2 $ and $\beta \to \beta + n_2 b_1 
+ m_2 b_2 $, 
we do a resummation over $b_1$ and $b_2$:
\begin{eqnarray}
V &  = & { 1\over 2^4} 
\sum_{\alpha, \beta \in \Xi_0}  \sum_{n, m, \tilde{n}, 
\tilde{m}=0 }^1
\sum_{n_1,n_2,m_1,m_2=0}^1
C\Big[  \begin{array}{c} \alpha + n b_3 + {m}b_4  +n_1
 b_1 + m_1 b_2 \cr
\beta  + \tilde{n} b_3 + \tilde{m}b_4  +n_2 b_1 + m_2 b_2
\end{array} \Big]
\nonumber \\
& & \quad \times  \prod_{f\in F'} \Theta^{1/2}  
\Big[  \begin{array}{c} (\alpha+ n b_3 + {m}b_4  +n_1 b_1
 + m_1 b_2 )(f) \cr
(\beta + \tilde{n} b_3 + \tilde{m}b_4  +n_2 b_1 + m_2 b_2) 
(f) \end{array} 
\Big] (\tau)
\nonumber \\
&\equiv  & {1 \over 2^4} \sum_{\alpha, \beta \in \Xi_0} 
 \sum_{n, m, \tilde{n}, 
\tilde{m}=0 }^1
V\Big[  \begin{array}{c} \alpha + n b_3 + {m}b_4  \cr
\beta  + \tilde{n} b_3 + \tilde{m}b_4  \end{array} \Big], 
\label{defvv}
\end{eqnarray}
defining $V[\,]$ in an obvious notation. It is easy to prove
 the following property 
for $V[\,]$:
\begin{equation}
V\Big[  \begin{array}{c} \alpha + n b_3 + {m}b_4  \cr
\beta  + \tilde{n} b_3 + \tilde{m}b_4  \end{array} \Big] = 
V\Big[  \begin{array}{c} \alpha + n b_3 + {m}b_4  +n_1 b_1 
+ m_1 b_2 \cr
\beta  + \tilde{n} b_3 + \tilde{m}b_4  +n_2 b_1 + m_2 b_2
\end{array} \Big], 
\end{equation}
for any $n_1,n_2, m_1, m_2=0,1$ and $\alpha, \beta \in 
\Xi_0$. Now we define a 
subset $\Xi_0'$ of $\Xi_0$: 
$\Xi_0' \equiv \{ \alpha \in \Xi_0| \alpha \cap b_1 = \alpha 
\cap b_2 = \emptyset \} 
= \{\emptyset, F+b_1 + b_2\} $. 
By using the the above property of $V$  we can restrict the 
summation over $\alpha$
 and $\beta$ to $\Xi_0'$:
\begin{equation}
V = \sum_{\alpha , \beta \in \Xi_0'} \sum_{n, m, \tilde{n}, 
\tilde{m}=0,1 }
V\Big[  \begin{array}{c} \alpha + n b_3 + {m}b_4  \cr
\beta  + \tilde{n} b_3 + \tilde{m}b_4  \end{array} \Big]. 
\end{equation}

In the following we will prove that all these $V[\,]$'s are 
identically zero.  Before 
doing this let us first classify all these
different $V[\,]$'s according to their modular transformation
property. In fact all these $V[\,]$'s  can be classified into five
 categories. They are:

1) $V\Big[  \begin{array}{c} \alpha \cr
\beta   \end{array} \Big]$

2) $V\Big[  \begin{array}{c} \alpha + b_3 \cr
\beta  \end{array} \Big]$,  
$V\Big[  \begin{array}{c} \alpha \cr
\beta + b_3  \end{array} \Big]$ and 
$V\Big[  \begin{array}{c} \alpha + b_3 \cr
\beta + b_3  \end{array} \Big]$;

3-4) obtained from the above by substitution $b_3 \to b_4$ or
  $b_3 \to b_3 + b_4$;

5) $V\Big[  \begin{array}{c} \alpha + b_3 \cr
\beta  +b_4 \end{array} \Big]$,
$V\Big[  \begin{array}{c} \alpha + b_3 \cr
\beta  + b_3+ b_4 \end{array} \Big]$,
$V\Big[  \begin{array}{c} \alpha + b_4 \cr
\beta  + b_3 + b_4 \end{array} \Big]$,
$V\Big[  \begin{array}{c} \alpha + b_4 \cr
\beta  +b_3 \end{array} \Big]$,

\noindent
$V\Big[  \begin{array}{c} \alpha + b_3 + b_4 \cr
\beta  +b_3 \end{array} \Big]$,
$V\Big[  \begin{array}{c} \alpha + b_3 + b_4 \cr
\beta  +b_4 \end{array} \Big]$.

All the $V[\,]$'s in the same category can be transformed into 
each other by modular 
transformations. This can easily be shown by making use of the
 following generating 
modular transformations at one-loop:
\begin{eqnarray}
\Theta \Big[  \begin{array}{c} a \cr
b    \end{array} \Big] (\tau) & \stackrel{\mathrm{S
}}{\longrightarrow} & 
\Theta \Big[  \begin{array}{c} b \cr
a   \end{array} \Big] (-1/\tau) ,
\\
\Theta \Big[  \begin{array}{c} a \cr
b    \end{array} \Big] (\tau) & \stackrel{\mathrm{
T}}{\longrightarrow}& 
\Theta \Big[  \begin{array}{c} a \cr
a  +b +1  \end{array} \Big] (\tau+ 1 ) .
\end{eqnarray}
The property of the coefficient $C[\,]$'s ensures that all
$V[\,]$'s transform nicely and we have
\begin{eqnarray}
V \Big[  \begin{array}{c} \alpha  \cr
\beta     \end{array} \Big] (\tau) & \stackrel{\mathrm{
S}}{\longrightarrow} & 
V \Big[  \begin{array}{c} \beta   \cr
\alpha   \end{array} \Big] (-1/\tau) ,
\\
V\Big[  \begin{array}{c} \alpha  \cr
\beta      \end{array} \Big] (\tau) & \stackrel{\mathrm{T}}{
\longrightarrow}& 
V \Big[  \begin{array}{c} \alpha  \cr
\alpha  +\beta  +b_0\end{array} \Big] (\tau+ 1 ) .
\end{eqnarray}
The following demonstrates how we transform
$V\Big[  \begin{array}{c} \alpha  + b_3 \cr
\beta  + b_4      \end{array} \Big] $ to 

\noindent
$V\Big[  \begin{array}{c} \tilde{\alpha}  + b_3 + b_4  \cr
\tilde{\beta}  + b_4     \end{array} \Big] $ by a series of
 modular transformations:
\begin{eqnarray}
V\Big[  \begin{array}{c} \alpha  + b_3 \cr
\beta  + b_4      \end{array} \Big](\tau)  
& \stackrel{\mathrm{S}}{\longrightarrow} & 
V\Big[  \begin{array}{c} \beta  + b_4 \cr
\alpha  + b_3 \end{array} \Big](-1/\tau)  
\nonumber \\
& \stackrel{\mathrm{T}}{\longrightarrow} & 
V\Big[  \begin{array}{c} \beta  + b_4 \cr
\alpha + \beta  + b_0 + b_3 + b_4      \end{array} \Big](-1/
\tau +1 )  
\nonumber \\
& \stackrel{\mathrm{S}}{\longrightarrow} & 
V\Big[  \begin{array}{c} \alpha + \beta  + b_0 + b_3 
+ b_4      \cr\beta  + b_4 
\end{array} \Big](\tau/(1-\tau) )
\nonumber \\
& = & V\Big[  \begin{array}{c} {\alpha} +\beta +b_0 + b_1 + b_2 + b_3 + b_4  \cr
{\beta}  + b_4     \end{array} \Big]  (\tau/(1-\tau) )  
\nonumber \\
&   \equiv & 
V\Big[  \begin{array}{c} \tilde{\alpha}  + b_3 + b_4  \cr
\tilde{\beta}  + b_4     \end{array} \Big]  (\tau/(1-\tau) )   ,
\end{eqnarray}
with $\tilde{\alpha} = \alpha  + \beta  + b_0 + b_1  + b_2 $ and $\tilde{
\beta } = \beta$. One 
easily ckecks that 
\begin{equation}
\alpha, \beta \in \Xi'_0 \Longrightarrow \tilde{\alpha } \in \Xi'_0.
\end{equation}

After this detour we can prove the vanishing of all $V[\,]$'s
in one category by just 
proving the vanishing of just one of the $V[\,]$ in this category.
 Modular transformation
 ensures the vanishing of all the other $V[\,]$'s because they are
 related with each other 
by modular transormations. 

The vanishing of $V[\,]$'s in the first 4 categories is automatic 
because of supersymmetry. 
On the other hand the vanishing of $V[\,]$'s  in the fifth categories is 
because there always exists 
an odd spin structure in the theta function product.  
Let us prove these assertions in the following.

To prove $V\Big[  \begin{array}{c} \alpha \cr
\beta   \end{array} \Big]$ =0, let us first recall the definition of 
it given in (\ref{defvv}):
\begin{equation}
V\Big[  \begin{array}{c} \alpha \cr
\beta   \end{array} \Big] = \sum_{n_i,m_i = 0}^1 C\Big[ 
 \begin{array}{c} \alpha  + n_1
 b_1+ m_1 b_2 \cr
\beta  + n_2 b_1 + m_2 b_2   \end{array} \Big] \prod_{f\in F'} 
\Theta^{1/2} 
 \Big[  \begin{array}{c} (\alpha  + n_1 b_1+ m_1 b_2)(f) \cr
(\beta  + n_2 b_1 + m_2 b_2)(f)   \end{array} \Big] .
\end{equation}
Now we separate the transverse fermions $F'$ into three 
non-intersecting  subsets $b_1$, 
$b_2 $ and $F' + b_1  + b_2$. Then we have\footnote{Remember
$\alpha \cap b_i = \beta\cap b_i = \emptyset, i = 1,2$.}
\begin{eqnarray}
V\Big[  \begin{array}{c} \alpha \cr
\beta   \end{array} \Big] & = & \sum_{n_i,m_i = 0}^1 
C\Big[  \begin{array}{c} \alpha 
 + n_1 b_1+ m_1 b_2 \cr
\beta  + n_2 b_1 + m_2 b_2   \end{array} \Big]  
\prod_{f\in b_1} \Theta^{1/2} 
 \Big[  \begin{array}{c}  n_1 \cr
n_2   \end{array} \Big] (\tau) \nonumber \\
& & \times \prod_{f\in b_2} 
\bar\Theta^{1/2} 
 \Big[  \begin{array}{c} m_1 \cr
m_2    \end{array} \Big] (\bar{\tau})
 \prod_{f\in F' + b_1 + b_2}
\Theta^{1/2} 
 \Big[  \begin{array}{c} \alpha(f) \cr
\beta(f)   \end{array} \Big] 
\nonumber \\
& = & \sum_{n_i,m_i = 0}^1 C\Big[  \begin{array}{c} 
\alpha  + n_1 b_1+ m_1 b_2 \cr
\beta  + n_2 b_1 + m_2 b_2   \end{array} \Big]  
 \Theta^{4} 
 \Big[  \begin{array}{c}  n_1 \cr
n_2   \end{array} \Big] (\tau)  \bar\Theta^4 
\Big[  \begin{array}{c} m_1 \cr
m_2    \end{array} \Big] (\bar{\tau})
\nonumber \\
& & \quad \times \prod_{f\in F' + b_1 + b_2}
\Theta^{1/2} 
 \Big[  \begin{array}{c} \alpha(f) \cr
\beta(f)   \end{array} \Big].
\label{defcc}
\end{eqnarray}
Noticing $\alpha\cap b_i= \beta\cap b_i = \emptyset$, one 
can derive the following:
\begin{eqnarray}
& & C\Big[  \begin{array}{c} \alpha  \cr
n b_1    \end{array} \Big]   = 
C\Big[  \begin{array}{c} \alpha  \cr
n b_2    \end{array} \Big]    = (-1)^n, 
\\
& & C\Big[  \begin{array}{c} \alpha  + n_1 b_1+ m_1 b_2 \cr
\beta  + n_2 b_1 + m_2 b_2   \end{array} \Big]   = (-1)^{ n_1
 + n_2 + m_1 + m_2 } \, 
C\Big[  \begin{array}{c} \alpha  \cr
\beta     \end{array} \Big] . 
\end{eqnarray}
By using these results in (\ref{defcc}) we have 
\begin{eqnarray}
V\Big[  \begin{array}{c} \alpha \cr
\beta   \end{array} \Big] & \propto & \sum_{n_i=0}^1 (-1)^{n_1
 + n_2} \, \Theta^4  
\Big[  \begin{array}{c}  n_1 \cr
n_2   \end{array} \Big] (\tau)  \sum_{m_i=0}^1 (-1)^{m_1 
+ m_2} \, \Theta^4  \Big[ 
 \begin{array}{c}  m_1 \cr
m_2   \end{array} \Big] (\bar{\tau})  
\nonumber \\
& = & \left( \Theta^4 \Big[  \begin{array}{c}  0 \cr
0   \end{array} \Big] (\tau)   -
\Theta^4 \Big[  \begin{array}{c}  0 \cr
1   \end{array} \Big] (\tau)   -
\Theta^4 \Big[  \begin{array}{c}  1 \cr
0   \end{array} \Big] (\tau)   +
\Theta^4 \Big[  \begin{array}{c}  1 \cr
1   \end{array} \Big] (\tau)   \right) \times (C. C.) 
\nonumber \\
&  = &  0, \label{complexconjugate}
\end{eqnarray}
by making use of the well-known identity:
\begin{equation}
\Theta^4 \Big[  \begin{array}{c}  0 \cr
0   \end{array} \Big] (\tau)   -
\Theta^4 \Big[  \begin{array}{c}  0 \cr
1   \end{array} \Big] (\tau)   -
\Theta^4 \Big[  \begin{array}{c}  1 \cr
0   \end{array} \Big] (\tau)    = 0, 
\end{equation}
and $\Theta \Big[  \begin{array}{c}  1 \cr
1   \end{array} \Big] (\tau)=0$. 
   In eq. (\ref{complexconjugate}),  ($C.C.$) denotes 
complex conjugation.
  
To prove  $V\Big[  \begin{array}{c} \alpha + b_3 \cr
\beta  \end{array} \Big] =0$, one derives
\begin{equation}
C\Big[  \begin{array}{c} \alpha + b_3  + n_1 b_1+ m_1 b_2 \cr
\beta   +  n_2 b_1 + m_2 b_2   \end{array} \Big]   = (-1)^{  n_1+ 
m_1 } 
\, C\Big[  \begin{array}{c} \alpha  + b_3 \cr
\beta     \end{array} \Big] 
\, C\Big[  \begin{array}{c} b_3    \cr
n_2 b_1       \end{array} \Big] 
\, C\Big[  \begin{array}{c} b_3    \cr
m_2 b_2      \end{array} \Big] , 
\end{equation}
and we have 
\begin{eqnarray}
V\Big[  \begin{array}{c} \alpha + b_3  \cr
\beta  \end{array} \Big]  & \propto & 
\sum_{n_i,m_i=0}^1 (-1)^{  n_1 + m_1 } 
\, C\Big[  \begin{array}{c} b_3    \cr
n_2 b_1       \end{array} \Big] 
\, C\Big[  \begin{array}{c} b_3    \cr
m_2 b_2      \end{array} \Big]   
\Theta^2 \Big[  \begin{array}{c}  n_1 \cr
n_2   \end{array} \Big] (\tau)  
\nonumber \\
& & \times \Theta^2 \Big[  \begin{array}{c}  n_1+1    \cr
n_2 \end{array} \Big] (\tau)
\Theta^4 \Big[  \begin{array}{c}  m_1 \cr
m_2   \end{array} \Big] (\bar{\tau})    
\nonumber \\
& \propto & \sum_{n_1 =0}^1 (-1)^{n_1} 
\Theta^2 \Big[  \begin{array}{c}  n_1  \cr
0   \end{array} \Big] (\tau)  
\Theta^2 \Big[  \begin{array}{c}  n_1 + 1  \cr
0   \end{array} \Big] (\tau)
\nonumber \\
& = & 
\Theta^2 \Big[  \begin{array}{c}  0 \cr
0   \end{array} \Big] (\tau)  
\Theta^2 \Big[  \begin{array}{c}  1 \cr
0   \end{array} \Big] (\tau) - 
\Theta^2 \Big[  \begin{array}{c}  1 \cr
0   \end{array} \Big] (\tau)  
\Theta^2 \Big[  \begin{array}{c}  0 \cr
0   \end{array} \Big] (\tau)  \nonumber \\
& = & 0.
\end{eqnarray} 
Here we have used the fact that because 
$\Theta \Big[  \begin{array}{c}  1 \cr 1  
\end{array} \Big] (\tau)=0$, only $n_2=0$ might give a
possibly non-vanishing  contribution to $V$. 

Finally the proof of  $V\Big[  \begin{array}{c} \alpha  + b_3 \cr
\beta  + b_4  \end{array} \Big] =0$ is based on the following
 results ($b\equiv F + b_1 + b_2 $):
\begin{eqnarray}
\alpha = \beta = \emptyset, \quad & &  
b_3\cap b_4 = \{ y^{I=1,\cdots, 4}, \tilde{y}^{I=1,\cdots,4} \},
\\
\alpha = \emptyset, \beta = b, \quad & &
b_3\cap ( \beta + b_4) = \{ \tilde{\omega}^{I=1,\cdots,4} \},
\\
\beta = \emptyset, \alpha = b , \quad & &
(\alpha + b_3)\cap  b_4 = \{ {\omega}^{I=1,\cdots,4} \},
\\
\alpha = \beta = b , \quad & &
(\alpha + b_3)\cap  (\beta +b_4) = \{ y^I, \tilde{y}^I, \omega^I,
 \tilde{\omega}^I| I = 5,6 \}.
\end{eqnarray}
By using these results we have 
\begin{eqnarray}
V\Big[  \begin{array}{c} \alpha  + b_3 \cr
\beta  + b_4  \end{array} \Big]  & \propto & 
\prod_{f\in F' + b_1 + b_2 } \Theta^{1/2} 
\Big[  \begin{array}{c} (\alpha  + b_3)(f) \cr
(\beta  + b_4)(f)  \end{array} \Big] 
\nonumber \\ 
& \propto & \Theta^{2} 
\Big[  \begin{array}{c}1 \cr
1  \end{array} \Big] (\tau) ~\hbox{or}~ \bar\Theta^{2} 
\Big[  \begin{array}{c}1 \cr
1  \end{array} \Big] (\bar{\tau})  = 0 .
\end{eqnarray}

In the next section we extend the above reasoning to 
two-loop computations, but we will find a non-vanishing result
 for the cosmological 
constant before the integration over the moduli space. 

\section{Two-loop cosmological constant}

As we recalled above, in two-loop computation another factor must be
inserted into the sum over the even spin structures, representing the
supercurrent correlator related to the occurrence of superghost zero modes
(the odd spin structures give a trivially zero contribution in this case too).
Its exact form is quite involved, however for checking the 
vanishing   of the
cosmological constant we only need few relevant facts, which we will
recall here. The computation of the 
two-loop cosmological constant
 (the left part) follows the standard strategy 
\cite{GavaIengoSotkov,Verlinde,VerlindeTrieste} of first 
integrating over the supermoduli by choosing
the super-Beltrami differentials to be 
$\delta$-functions in moduli independent 
points $x_{1,2}$ on the Riemann surface. One thus obtains for the left
part of the integrand (to be integrated over the moduli of the surface,
represented by three of the branch points $a_i$, the other three being
fixed by the $SL(2,C)$ projective invariance, see below):
 \begin{equation}
V_L = (\det \bar{\partial}_1)^{-D/2} \det \bar{\partial}_2
 (\det \bar{\partial}_{3/2}
)^{-1}_{s(\beta,\gamma)} 
{ \langle J(x_1) J(x_2) \rangle_{s(\psi)} \over \det 
\Phi_a^{3/2}(x_b) } \, \prod_{f\in F_{
\mathrm{L}} }
 (\det \bar{\partial}_{1/2} )^{1/2}_{s(f)} ,  \label{vll}
\end{equation}
where $\det \bar{\partial}_j$ stands for the usual 
chiral determinant and $\Phi_a^{3/2}(x), 
a=1,2$ are the holomorphic $3/2$-differentials. 
All the determinants, the correlators 
and $3/2$-differentials appearing in eq. (\ref{vll}) 
can be explicitly expressed in the 
hyperelliptic description, where the genus $g=2$ 
Riemann surface is represented as 
lower and upper (compactified) sheets of the 
function
\begin{equation}
y^2(z) = \prod_{i=1}^6 (z-a_i),
\end{equation}
and the branch points $a_i$ define three branch cuts in 
each sheet. Three $a_i$ can be 
fixed by $SL(2,C)$ projective invariance and the other three are
 taken to be moduli. The two 
independent 
holomorphic one-forms $w_i(z)$ can be taken to be 
\begin{equation}
w_1(z) = { 1\over y(z) } , \qquad \qquad 
w_2(z) = { z\over y(z) }. 
\end{equation}
As in ref. \cite{GavaIengoSotkov} we choose the positions 
$x_{1,2}$ of the 
super-Beltrami differentials to be the two simple zeros of $w_1$,
 i.e. $x_1 = \infty$ 
in the upper sheet and $x_2 = \infty$ in the lower sheet 
respectively. In this language 
the even spin structures are equivalent  to ten different
 splitting of the six branch points 
$\{a_i\}, i=1,\cdots, 6$ into two non-ordered sets $\{A_i\}, 
\{B_i\}, i=1,2,3$. 

With a convenient normalization 
\cite{GavaIengoSotkov}  
we have $\det \Phi^{3/2}_a(x_b)=1$, and the spin-structure 
dependent part of 
$ \langle J(x_1) J(x_2) \rangle $
 is carried by  the expressions 
\begin{eqnarray}
Y_1[s(\psi)] & \equiv  & 2\, \langle \psi(x_1)\psi(x_2) 
\rangle_{s(\psi)} = { 1\over 2} 
\sum_{i=1}^3 (A_i - B_i), 
\\
{Y}_2[s(\psi)] & \equiv  & 4\, \langle \partial \psi(x_1)
\psi(x_2) \rangle_{s(\psi)} = 
{ 1\over 2} \sum_{i=1}^3 (A_i^2- B_i^2), 
\end{eqnarray}
coming from the $J_{\psi} =  \psi \cdot \partial X $ 
and $J_{\mathrm{gh}}$ part 
of eq. (\ref{eqone}) (see ref. \cite{GavaIengoSotkov} 
 for details). Similarly, the 
compactified sector contribution is 
\begin{equation}
\langle J_{\chi}(x_1) J_{\chi}(x_2) \rangle_{s(\psi)} = 
- \sum_{I=1}^6 Y_1[s(\chi^I)] \, Y_1[s(y^I)] \, Y_1
[ s(\omega^I)]. 
\end{equation}
With all these expressions in hand we can write 
\begin{equation}
{ \langle J(x_1) J(x_2) \rangle_{s(\psi)} } = 
K_1 \, Y_1[ s(\psi)] + K_2 \, {Y}_2[ s(\psi)] -  
\sum_{I=1}^6 Y_1[s(\chi^I)] \, 
Y_1[s(y^I)] \, Y_1[ s(\omega^I)], 
\end{equation}
where $K_i$'s are spin structure independent (to be 
precise $K_i$'s are symmetric
 for every $a_i \leftrightarrow a_j$ interchange). 

By using the above hyperelliptic representation of the 
genus 2 Riemann surface, 
one can compute all the determinants in (\ref{vll}) 
explicityly and we have the 
following result for $V_L$:
\begin{equation}
V_L = {1 \over (\det K)^{ D/2} \prod_{i<j} a_{ij} }  \, 
{ \langle J(r+) J(r-) \rangle_{s(\psi)} }
\prod_{f\in F_L'} Q_{s(f)}^{1/8} , 
\end{equation}
where we have put $x_1 =r+$ and $x_2=r-$, meaning
 two arbitrary points (not 
necessarily at $\infty$) on the lower and upper sheets 
of the complex plane. Also the quantity $Q_s$  is proportional
(with a spin structure independent factor), by means
of the Thomae formula \cite{Fay} to $\theta_s^4 (0)$.
In terms of the above sets $\{A_i\},\{B_i\}$ we have:
\begin{equation}
Q_s=\prod^3_{i<j} (A_i-A_j)(B_i-B_j).  \label{QQ}
\end{equation}   
By  using this result we can write down the two-loop 
cosmological constant as follows:
\begin{equation}
\Lambda = \int  { {d^2 \mu} \over T^{D/2}
 \prod_{i<j} |a_{ij}|^2 } 
 \sum_{\alpha,\beta \in \Xi} 
C \Big[  \begin{array}{c} \alpha\cr
\beta\end{array} \Big]
{ \langle J(r+) J(r-) \tilde{J}(\bar{s}+) \tilde{J}(\bar{s}-)
 \rangle_{s(\psi,\tilde{\psi})} }
\prod_{f\in F'} Q_{s(f)}^{1/8}, 
\end{equation}
where $d^2 \mu = { \prod_{i=1}^6 d^2 a_i \over
 \prod_{i<j} |a_{ij}|^2 \, d V}$ is
 the measure of the two-loop moduli space and 
$ d V = { d^2 a_i d^2 a_j d^2 a_k 
\over |a_{ij}a_{jk}a_{ki}|^2 }$ is the infinitesimal 
volume of the $SL(2,C)$
 tranformation. 

In type II superstring the correlator $\langle JJ\tilde J\tilde J\rangle$
is not completely left-right factorized. However as far as
its spin structure dependence is concerned (i.e. what is relevant to check
the zero of the cosmological constant) it is a sum of factorized pieces.
Therefore, since we will check separately the terms containing 
$Y_1$ and $Y_2$, we can assume the supercurrent correlator,
denoted for short $X_{long,3/2}$, to be effectively
factorized:
$X_{long,3/2} \equiv  \langle JJ\tilde J\tilde J\rangle = 
\langle JJ\rangle \langle \tilde J\tilde J \rangle $. Only when we want
the complete expression for the cosmological constant, we should take into
account this left-right coupling.

\section*{\centerline{Organization of the computation}}

At two loops one can do a resummation over $b_1$ and 
$b_2$ just as we did at 
one loop order, by making use of the same subsets $\Xi_0$ and $\Xi_0'$
introduced at one loop. We
have  {
\arraycolsep=3pt
\begin{eqnarray}
V_{{Two-loops}} & = & \sum_{\alpha_i,\beta_i \in \Xi} X_{long,3/2} C
\Big[  \begin{array}{cc} \alpha_1 & \alpha_2\cr
\beta_1 & \beta_2   \end{array} \Big]
\, \prod_{f\in F'} Q^{1/8} 
\Big[  \begin{array}{cc} \alpha_1(f) & \alpha_2(f) \cr
\beta_1 (f) & \beta_2(f)    \end{array} \Big] 
\nonumber \\
& = & \sum_{\alpha_i,\beta_i \in \Xi_0} \sum_{n_i,m_i,
\tilde{n}_i,\tilde{m}_i=0}^1 
X_{long,3/2} C
\Big[  \begin{array}{cc} \alpha_1 + n_1 b_3 + m_1 b_4 & 
\alpha_2 + n_2 b_3 
+ m_2 b_4 \cr
\beta_1+ \tilde{n}_1 b_3 + \tilde{m}_1 b_4  & \beta_2 +
 \tilde{  n}_2 b_3 + 
\tilde{m}_2 b_4  \end{array} \Big]
\nonumber \\
&  & \times 
\, \prod_{f\in F'} Q^{1/8} 
\Big[  \begin{array}{cc} (\alpha_1 + n_1 b_3 + m_1 b_4 )
 (f) & (\alpha_2+ 
n_2 b_3 + m_2 b_4 ) ( f) \cr
(\beta_1+ \tilde{n}_1 b_3 + \tilde{m}_1 b_4  )
 (f) &(  \beta_2+ \tilde{  n}_2 b_3 + \tilde{m}_2 b_4  ) (f)    
\end{array} \Big] 
\nonumber \\
&  = & \sum_{\alpha_i,\beta_i \in \Xi'_0} \sum_{n_i,m_i,
\tilde{n}_i,\tilde{m}_i=0}^1 
V
\Big[  \begin{array}{cc} \alpha_1 + n_1 b_3 + m_1 b_4 & 
\alpha_2 + n_2 b_3 
+ m_2 b_4 \cr
\beta_1+ \tilde{n}_1 b_3 + \tilde{m}_1 b_4  & \beta_2 +
 \tilde{  n}_2 b_3 +
 \tilde{m}_2 b_4  \end{array} \Big], 
\end{eqnarray}
where }
\begin{eqnarray}
V\Big[  \begin{array}{cc} \alpha_1 & \alpha_2\cr
\beta_1 & \beta_2   \end{array} \Big]  \equiv   \sum_{n_i,m_i,
\tilde{n}_i,
\tilde{m}_i=0}^1 
X_{long,3/2} C
\Big[  \begin{array}{cc} \alpha_1 + n_1 b_1 + m_1 b_2 &
 \alpha_2 + n_2 b_1 
+ m_2 b_2\cr
\beta_1+ \tilde{n}_1 b_1 + \tilde{m}_1 b_2  & \beta_2 +
 \tilde{  n}_2 b_1 
+ \tilde{m}_2 b_2  \end{array} \Big]  & &  
\nonumber \\
 \times 
\, \prod_{f\in F'} Q^{1/8} 
\Big[  \begin{array}{cc} (\alpha_1 + n_1 b_1 + m_1 b_2 ) (f) 
& (\alpha_2+ n_2 b_1
 + m_2 b_2 ) ( f) \cr
(\beta_1+ \tilde{n}_1 b_1 + \tilde{m}_1 b_2  )
 (f) &(  \beta_2+ \tilde{  n}_2 b_1 + \tilde{m}_2 b_2  ) (f)    
\end{array} \Big] .  
 \quad  \quad \quad \nonumber 
\end{eqnarray}

As in one-loop computation we classify all the $V[\,]$'s into 
6 categories (see Appendix B). They
 are ($\alpha_i, \beta_i \in \Xi'_0$):

1) $V\Big[  \begin{array}{cc} \alpha_1 & \alpha_2\cr
\beta_1 & \beta_2   \end{array} \Big]$;

2) $V\Big[  \begin{array}{cc} \alpha_1 + b_3 &  \alpha_2 \cr
\beta_1 & \beta_2   \end{array} \Big]$ and 14 other $V[\,]$'s 
obtained by 
two-loop modular transformations;

3-4) obtained from the above by substitution $b_3 \to b_4$ or 
$b_3 \to b_3 + b_4$;

5) $V\Big[  \begin{array}{cc} \alpha_1 + b_3 &  \alpha_2 \cr
\beta_1 + b_4 & \beta_2   \end{array} \Big]$ and 119 other
$V[\,]$'s obtained  by two-loop modular transformations;

6) $V\Big[  \begin{array}{cc} \alpha_1 + b_3 
&  \alpha_2  + b_4 \cr
\beta_1  & \beta_2   \end{array} \Big]$ and 89 other $V[\,]$'s 
obtained by  two-loop modular transformations.

The total number of different $V[\,]$'s (not counting different
 $\alpha_i,  \beta_j\in \Xi_0'$) is $(2^2)^4 = 256$
which is the right  number of different 
$V[\,]$'s in the above classification.

\section*{\centerline{Computing the various terms}}

\subsection{The susy preserving terms.} 

The $V[\,]$'s in the first 4 categories
are zero  because of supersymmetry. This was 
proved rigorously in \cite{IengoGalenI}. We 
will give the proof in the Appendix C. 
This is a useful  exercise because we will see quite clearly  
the difference between the supersymmetric 
model and the non-supersymmetric  model. 

We note that this proof makes  
 use of the following identities ($i=1,2$):
\begin{eqnarray}
 & & \left( Y_i \Big[  \begin{array}{cc} 0 & 0 \cr
0  &  0    \end{array} \Big]  -
Y_i    \Big[  \begin{array}{cc} 1 & 0 \cr
0  &  0    \end{array} \Big]  \right)
Q^{1/2} \Big[  \begin{array}{cc} 0 & 0 \cr
0  &  0    \end{array} \Big]  \,
Q^{1/2} \Big[  \begin{array}{cc} 1 & 0 \cr
0  &  0    \end{array} \Big]
\nonumber \\  
& & 
\quad -\left( Y_i \Big[  \begin{array}{cc} 0 & 1 \cr
0  &  0    \end{array} \Big]  -
Y_i    \Big[  \begin{array}{cc} 1 & 1 \cr
0  &  0    \end{array} \Big]  \right)
Q^{1/2} \Big[  \begin{array}{cc} 0 & 1 \cr
0  &  0    \end{array} \Big]  \,
Q^{1/2} \Big[  \begin{array}{cc} 1 & 1 \cr
0  &  0  \end{array} \Big]      
\nonumber \\
& & \quad -\left( Y_i \Big[  \begin{array}{cc} 0 & 0 \cr
0  &  1    \end{array} \Big]  -
Y_i    \Big[  \begin{array}{cc} 1 & 0 \cr
0  &  1    \end{array} \Big]  \right)
Q^{1/2} \Big[  \begin{array}{cc} 0 & 0 \cr
0  &  1    \end{array} \Big]  \,
Q^{1/2} \Big[  \begin{array}{cc} 1 & 0 \cr
0  &  1    \end{array} \Big]= 0,
\label{diffaa}
\end{eqnarray}
which was conjectured in \cite{ABKb} and proved explicitly in \cite{IengoGalenI}.

\subsection{The terms containg an odd Theta-function.}

Now we prove the vanishing of
$V\Big[  \begin{array}{cc} \alpha_1 +  b_3 & \alpha_2 \cr
\beta_1 + b_4 & \beta_2   \end{array} \Big]  $. We have
\begin{equation}
V\Big[  \begin{array}{cc} \alpha_1 +  b_3 & \alpha_2 \cr
\beta_1 + b_4 & \beta_2   \end{array} \Big]   \propto
\prod_{f \in F+b_1 + b_2 } \Theta^{1/2}
\Big[  \begin{array}{cc} (\alpha_1 +  b_3)(f) & \alpha_2(f) \cr
(\beta_1 + b_4)(f) & \beta_2(f)   \end{array} \Big]  .
\end{equation}
The strategy to prove the vanishing of the above expression is to
show that there always exist at least one odd spin structure for the
$\Theta$-function. As we proved in section 2, for any choice of
$ \alpha_1$ and $\beta_1$, one can find fermionic field $f$ in
$b=F+ b_1 + b_2$ such that
$ \Big[  \begin{array}{c} \alpha_1 +  b_3 \cr
\beta_1 + b_4 \end{array} \Big]   =
\Big[  \begin{array}{c} 1\cr
1 \end{array} \Big]$.  For such $f$,
$\Big[  \begin{array}{cc} (\alpha_1 +  b_3)(f) & \alpha_2(f) \cr
(\beta_1 + b_4)(f) & \beta_2(f)   \end{array} \Big]  $ takes the 
 values
$\Big[  \begin{array}{cc} 1 & 0 \cr
1 & 0   \end{array} \Big]  $,
$\Big[  \begin{array}{cc} 1 & 1 \cr
1 & 0   \end{array} \Big]  $ or
$\Big[  \begin{array}{cc} 1 & 0 \cr
1 & 1   \end{array} \Big]  $ for
$ \Big[  \begin{array}{c} \alpha_2 \cr
\beta_2 \end{array} \Big]   =
\Big[  \begin{array}{c} \emptyset \cr
\emptyset \end{array} \Big]$,
$\Big[  \begin{array}{c} b  \cr
\emptyset \end{array} \Big]$ or
$\Big[  \begin{array}{c} \emptyset \cr
b \end{array} \Big]$ respectively.
All these are  odd spin structures. For
$ \Big[  \begin{array}{c} \alpha_2 \cr
\beta_2 \end{array} \Big]   =
\Big[  \begin{array}{c} b   \cr
b   \end{array} \Big]$,  we have
\begin{equation}
\Big[  \begin{array}{cc} (\alpha_1 +  b_3)(f) & \alpha_2(f) \cr
(\beta_1 + b_4)(f) & \beta_2(f)   \end{array} \Big]    =
\Big[  \begin{array}{cc} (\alpha_1 +  b_3)(f) & 1 \cr
(\beta_1 + b_4)(f) & 1   \end{array} \Big]  .  
\end{equation}
To have a possibly non-vansihing result, this must be an even
spin structure for all $f\in b $. This requires
\begin{equation}
(\alpha_1 + b_3 )(f) = (\beta_1 + b_4 )(f) = 1, \forall f \in b,
\end{equation}
In particular this implies
\begin{equation}
(\alpha_1 + b_3 )\cap (b) = b,
\end{equation}
which is not the case for any choice of $\alpha_1 \in \Xi_0' = \{
\emptyset, b\}$.

\subsection{The last step: the non-vanishing terms}

The last step is to compute
$V\Big[  \begin{array}{cc}
 \alpha_1 +  b_3 & \alpha_2  + b_4 \cr
\beta_1 & \beta_2   \end{array} \Big]$. First we note the
following
result:
\begin{equation}
V\Big[  \begin{array}{cc}
 \alpha_1 +  b_3 & \alpha_2  + b_4 \cr
\beta_1 & \beta_2   \end{array} \Big] = 0,  \quad \hbox{for}
\quad \beta_1 = F+ b_1 + b_2\quad \hbox{or} \quad
\beta_2= F + b_1 + b_2,   
\end{equation}
which can be proved by following the method of the
last paragraph.
Now we compute
$V\Big[  \begin{array}{cc}
 \alpha_1 +  b_3 & \alpha_2  + b_4 \cr
\emptyset  & \emptyset   \end{array} \Big] $. For simplicity
let us  first take $\alpha_1 = \alpha_2 = \emptyset$ and
we have
\begin{eqnarray}
& & V\Big[  \begin{array}{cc}
 b_3 &  b_4 \cr
\emptyset  & \emptyset   \end{array} \Big]
=  \sum_{n_i,m_i,\tilde{n}_i,\tilde{m}_i=0}^1
\langle J(r+) J(r-) \tilde{J}(\bar{s}+)\tilde{J}(\bar{s}+)
\rangle_{s(\psi,\tilde{\psi})}
\nonumber \\
& & \quad \times
C\Big[  \begin{array}{cc}  b_3
+n_1 b_1 + m_1 b_2 &b_4 + n_2 b_1 + m_2 b_2 \cr
\tilde{n}_1 b_1  + \tilde{m}_1 b_2 &
\tilde{n}_2 b_1  + \tilde{m}_2 b_2 \end{array} \Big]
\nonumber \\
& & \quad \times
Q^{1/2} \Big[  \begin{array}{cc} n_1  & n_2 \cr
\tilde{n}_1  & \tilde{n}_2  \end{array} \Big] \,
Q^{1/2} \Big[  \begin{array}{cc} n_1 +1  & n_2 \cr
\tilde{n}_1  & \tilde{n}_2  \end{array} \Big] \,
\bar{Q}^{1/2} \Big[  \begin{array}{cc} m_1  & m_2 \cr
\tilde{m}_1  & \tilde{m}_2  \end{array} \Big]  
\bar{Q}^{1/2} \Big[  \begin{array}{cc} m_1  & m_2 +1 \cr
\tilde{m}_1  & \tilde{m}_2  \end{array} \Big]
\nonumber \\
& & \quad \times \left|
Q \Big[  \begin{array}{cc}0  & 0 \cr
0 & 0  \end{array} \Big] \,
\bar{Q} \Big[  \begin{array}{cc} 1  & 1 \cr
0 & 0  \end{array} \Big]
\right|
\, Q^{1/2} \Big[  \begin{array}{cc} 0  & 1 \cr 
0  & 0 \end{array} \Big] \,
\bar{Q}^{1/2} \Big[  \begin{array}{cc} 1  & 0 \cr
0  & 0 \end{array} \Big]  .
\label{sumcc}
\end{eqnarray}
The coeffieints $C$ we needed are given as follows:
\begin{eqnarray}
& & \hskip -1cm C\Big[  \begin{array}{cc}  b_3
+n_1 b_1 + m_1 b_2 &b_4 + n_2 b_1 + m_2 b_2 \cr
\tilde{n}_1 b_1  + \tilde{m}_1 b_2 &
\tilde{n}_2 b_1  + \tilde{m}_2 b_2 \end{array} \Big]
\nonumber \\
& & =
C\Big[  \begin{array}{c}  b_3
+n_1 b_1 + m_1 b_2 \cr
\tilde{n}_1 b_1  + \tilde{m}_1 b_2 \end{array} \Big]
\,C\Big[  \begin{array}{c}  b_4 + n_2 b_1 + m_2 b_2 \cr
\tilde{n}_2 b_1  + \tilde{m}_2 b_2 \end{array} \Big]
\nonumber \\
& & = C\Big[  \begin{array}{c}  b_3 
+n_1 b_1 \cr \tilde{n}_1 b_1 \end{array} \Big] \,   
C\Big[  \begin{array}{c}  b_3
+m_1 b_2 \cr \tilde{m}_1 b_2\end{array} \Big] \,
C\Big[  \begin{array}{c}  b_4
+n_2 b_1 \cr \tilde{n}_2 b_1 \end{array} \Big] \,
C\Big[  \begin{array}{c}  b_4
+m_2 b_2 \cr \tilde{m}_2 b_2\end{array} \Big] .
\label{coefficientsaa}
\end{eqnarray}
We note that in eq. (\ref{sumcc}) only $\tilde{n}_1 =
\tilde{m}_2 =0$
contribute. With this restriction the ambiguity in
eq. (\ref{coefficientsaa}) drops and we have
\begin{equation}
C\Big[  \begin{array}{c}  b_3
+n_1 b_1 \cr \tilde{n}_1 b_1 \end{array} \Big] = (-1)^{n_1},
\qquad C\Big[  \begin{array}{c}  b_4
+m_2 b_2 \cr \tilde{m}_2 b_2\end{array} \Big]  = (-1)^{m_2}.
\end{equation}
and
\begin{equation}
C\Big[  \begin{array}{c}  b_3
+m_1 b_2 \cr \tilde{m}_1 b_2\end{array} \Big] = (-1)^{m_1},
\quad
C\Big[  \begin{array}{c}  b_4
+n_2 b_1 \cr \tilde{n}_2 b_1 \end{array} \Big] =(-1)^{n_2}. 
\label{coefficientsbb}
\end{equation}
We note here that if we have (see eq. (\ref{coefficient}))
\begin{equation}
C\Big[  \begin{array}{c} b_1 \cr
b_4 \end{array} \Big] =-1 , \qquad \hbox{and} \qquad
C\Big[  \begin{array}{c}
b_2 \cr b_3 \end{array} \Big] =-1,
\end{equation}
eq. (\ref{coefficientsbb}) would become
\begin{equation}
C\Big[  \begin{array}{c}  b_3
+m_1 b_2 \cr \tilde{m}_1 b_2\end{array} \Big] = (-1)^{m_1 
+ \tilde{m}_1 },\quad
C\Big[  \begin{array}{c}  b_4   
+n_2 b_1 \cr \tilde{n}_2 b_1 \end{array} \Big] =(-1)^{n_2
+ \tilde{n}_2}.
\label{coefficientcc}
\end{equation}
Substituting eqs. (\ref{coefficientsaa})--(\ref{coefficientsbb}) into
eq. (\ref{sumcc}), the left part of spin structure dependent terms are
\begin{eqnarray}
 & &V_i =  \left( Y_i \Big[  \begin{array}{cc} 0 & 0 \cr
0  &  0    \end{array} \Big]  -
Y_i    \Big[  \begin{array}{cc} 1 & 0 \cr
0  &  0    \end{array} \Big]  \right)
Q^{1/2} \Big[  \begin{array}{cc} 0 & 0 \cr
0  &  0    \end{array} \Big]  \,
Q^{1/2} \Big[  \begin{array}{cc} 1 & 0 \cr
0  &  0    \end{array} \Big]
\nonumber \\
& & \quad
- \left( Y_i \Big[  \begin{array}{cc} 0 & 1 \cr
0  &  0    \end{array} \Big]  -
Y_i    \Big[  \begin{array}{cc} 1 & 1 \cr
0  &  0    \end{array} \Big]  \right)
Q^{1/2} \Big[  \begin{array}{cc} 0 & 1 \cr
0  &  0    \end{array} \Big]  \,
Q^{1/2} \Big[  \begin{array}{cc} 1 & 1\cr 
0  &  0    \end{array} \Big]
\nonumber \\
& & \quad +\left( Y_i \Big[  \begin{array}{cc} 0 & 0 \cr
0  &  1    \end{array} \Big]  -
Y_i    \Big[  \begin{array}{cc} 1 & 0 \cr
0  &  1    \end{array} \Big]  \right)
Q^{1/2} \Big[  \begin{array}{cc} 0 & 0 \cr
0  &  1    \end{array} \Big]  \,
Q^{1/2} \Big[  \begin{array}{cc} 1 & 0 \cr
0  &  1    \end{array} \Big] .
\label{diffbb}
\end{eqnarray}
We draw the attention of the readers to the difference between
eq. (\ref{diffaa}) and eq. (\ref{diffbb}). It's the minor (but crucial)
sign difference on the third line which gives a non-vanishing result.
We have
\begin{eqnarray}
V_i  & = & 2 \, \left( Y_i \Big[  \begin{array}{cc} 0 & 0 \cr
0  &  1    \end{array} \Big]  -
Y_i    \Big[  \begin{array}{cc} 1 & 0 \cr
0  &  1    \end{array} \Big]  \right)
Q^{1/2} \Big[  \begin{array}{cc} 0 & 0 \cr
0  &  1    \end{array} \Big]  \,
Q^{1/2} \Big[  \begin{array}{cc} 1 & 0 \cr
0  &  1    \end{array} \Big]
\nonumber \\
& = &  (a_2^i - a_3^i) \,
Q^{1/2} \Big[  \begin{array}{cc} 0 & 0 \cr
0  &  1    \end{array} \Big]  \,
Q^{1/2} \Big[  \begin{array}{cc} 1 & 0 \cr
0  &  1    \end{array} \Big] ,
\end{eqnarray}
if we set $r=\infty$ (the left supercurrent insertion points).
Similar calculation can be done also for the right part. The result is
\begin{equation}
\bar{V}_i = (\bar{a}_4^i - \bar{a}_5^i) \,
\bar{Q}^{1/2} \Big[  \begin{array}{cc} 0 & 0 \cr
1  &  0    \end{array} \Big]  \,
\bar{Q}^{1/2} \Big[  \begin{array}{cc} 0 & 1 \cr
1  &  0    \end{array} \Big] ,
\end{equation}
if we set $r=\infty$ (the right supercurrent insertion points).

By combining the left and right part, the  result is
\begin{equation}
V_{i\bar{j}}\Big[  \begin{array}{cc} b_3 & b_4 \cr
 \emptyset  &  \emptyset \end{array} \Big] =
a^i_{23}\bar{a}^j_{45} (a_{134526}\bar{a}_{164325})
|(124|356)^2(135|246)|,
\label{compaa}
\end{equation}
where $a_{ijklmn} \equiv a_{ij}a_{jk}a_{kl}a_{lm}a_{mn}
a_{ni}$ and similarly for $\bar{a}_{ijklmn}$. Here we have used the 
explicit expression for $Q$ eq.(\ref{QQ} ) and droped all other
non-essential factors. 

{\arraycolsep=3pt
For other $\alpha_i$, we can do similar calculation and the results are:
\begin{eqnarray}
V_{i\bar j}\Big[  \begin{array}{cc} b+  b_3 & b_4 \cr
 \emptyset  &  \emptyset \end{array} \Big] & = &
a^i_{23}\bar{a}^j_{45} (a_{124536}\bar{a}_{164235})
|(134|256)^2(125|346)|,
\label{compbb}
\\
V_{i\bar j}\Big[  \begin{array}{cc} b_3 & b+ b_4 \cr
 \emptyset  &  \emptyset \end{array} \Big] &  = &
a^i_{23}\bar{a}^j_{45} (a_{135426}\bar{a}_{165324})
|(125|346)^2(134|256)|,
\label{compcc}
\\
V_{i\bar j}\Big[  \begin{array}{cc} b+ b_3 & b+ b_4 \cr
 \emptyset  &  \emptyset \end{array} \Big] & = &
a^i_{23}\bar{a}^j_{45} (a_{125436}\bar{a}_{165234})
|(135|246)^2(124|356)| .
\label{compdd}
\end{eqnarray}
}

{From} the above results, we observe a general pattern for the
non-vanishing amplitude. By comparing eq. (\ref{compaa}) with 
eq. (\ref{compbb}) we see that  
$V_{i\bar j}\Big[  \begin{array}{cc} b+  b_3 & b_4 \cr
 \emptyset  &  \emptyset \end{array} \Big] $
can be obtained from
$V_{i\bar j}\Big[  \begin{array}{cc} b_3 & b_4 \cr
 \emptyset  &  \emptyset \end{array} \Big] $
by a modular transformation $a_2 \leftrightarrow a_3 $.
We also note that under this transformation
$\bar{a}_{164325}$ changes to $ - \bar{a}_{164235}$.
There is one other sign change from the left part: $a^i_{23} \to
-a^i_{23}$. We claim that all the other non-vanishing amplitude
can all be obtained from this single 
$V_{i\bar j}\Big[  \begin{array}{cc} b_3 & b_4 \cr
 \emptyset  &  \emptyset \end{array} \Big] $ non-vanishing expression by 
doing all the possible different modular transformations. To prove this claim
we introduce the following definition:
\begin{eqnarray}
 & & \hskip -1cm V_{i \bar{j}}(A_1,A_2,A_3,A_4,A_5,A_6) 
\nonumber \\
& & \equiv  (A^i_2-A^i_5)(\bar A_3^j-\bar A_4^j) (A_1A_2 A_3A_4 A_5A_6)  
(\bar{A}_1 \bar{A}_6\bar{A}_3
\bar{A}_2\bar{A}_5\bar{A}_4) 
\nonumber \\
& & \quad \times |(A_1 A_3 A_5|A_2 A_4 A_6)^2
(A_1 A_2 A_4 | A_3 A_5 A_6)|  .
\end{eqnarray}
where $(A_1,A_2,A_3,A_4,A_5,A_6)$ is a permutation of 
$(a_1,a_2,a_3,a_4,a_5,a_6)$ and $(A_1A_2 A_3A_4 A_5A_6)\equiv
(A_1-A_2)(A_2-A_3)(A_3-A_4)(A_4-A_5)(A_5-A_6)(A_6-A_1)$, etc.
It is easy to prove the following 
property of $V_i$:
\begin{equation}
V_{i\bar{j}}(A_6,A_5,A_4,A_3,A_2,A_1) = V_{i\bar{j}}
(A_1,A_2,A_3,A_4,A_5,A_6).
\end{equation}
By using this property, the number of different $V_{i\bar{j}}$
is $6!/2=360$. 
This is exactly the number of non-vanishing $V[\,]$'s: $90\times 4 = 360$
because every $V[\,]$ from the 6th category (there are 90 different of them
in this category) would give 4 non-vanishing terms as we explicitly showed in
eqs. (\ref{compaa})--(\ref{compdd}). 

By summing over all these permutations $\sigma$ we finally get the following
non-vanishing and explicitly modular invariant expression
\begin{equation}
V_{i\bar j} = \sum_{\sigma} V_{i\bar j}( 
\sigma(a_1), \sigma(a_2), \sigma(a_3), 
\sigma(a_4), \sigma(a_5), \sigma(a_6) ),  \label{ModularInvariant}
\end{equation}
which gives contribution to the integrand of the two-loop cosmological
constant.  We see no reason why a non-vanishng integrand should give a vanishing
cosmological constant after the integration over the moduli space.

\section*{Appendix A: Analysis of the gravitini content of the model}

Here we show that in the model of eq. (\ref{eq18}) there are no gravitini
in the free spectrum. The analysis is done by looking at the one-loop partition
function.  First, note that 
\begin{equation}
{ \mathrm{Tr}}\, (-)^{\beta}R^{\alpha}NS^{F+\alpha} \sim
\prod_{f\in F} \Theta^{1/2}
\Big[  \begin{array}{c}
\alpha (f)\cr 
\beta (f)\end{array} \Big],
\end{equation}
with a self-explanatory notation.
Therefore the partition function can be written as 
\begin{equation}
Z \sim \sum_{\alpha , \beta} C\Big[  \begin{array}{c}
\alpha \cr 
\beta \end{array} \Big]
{ \mathrm{Tr}}\,(-)^{\beta}R^{\alpha}NS^{F+\alpha}.
\end{equation}
Now by writing $\beta =\tilde\beta +n_ib_i$ with $n_i=0,1$   
and using eq. (\ref{setb}) and denoting 
$\delta_{\alpha}\equiv (-)^{\alpha (\psi ,\tilde\psi)}$,
we have
\begin{equation}
Z \sim \sum_{\alpha , \tilde\beta} 
{ \mathrm{Tr}}\,(-)^{\tilde\beta}
C\Big[  \begin{array}{c}
\alpha \cr  
\tilde\beta \end{array} \Big]
\Big( 1+ \delta_{\alpha}(-)^{b_i}C\Big[  \begin{array}{c}
\alpha \cr  
b_i \end{array} \Big] \Big)
R^{\alpha}NS^{F+\alpha}.
\end{equation}
Of course now $\tilde\beta$ runs over the set else than $b_i$.
We iterate this procedure for other base elements $b_j$ until
\begin{equation}
C\Big[  \begin{array}{c}
\alpha \cr 
\tilde\beta \end{array} \Big]
\to C\Big[  \begin{array}{c}
\alpha \cr 
\emptyset \end{array} \Big] =\delta_{\alpha}, 
\end{equation}
(the last equality follows from eq. (\ref{setb})), and finally we get
the following projections:
\begin{equation}
Z \sim \sum_{\alpha}  \delta_{\alpha}\prod_{i=0}^N 
\frac{1}{2}\Big( 1+\delta_{\alpha} (-)^{b_i}
C\Big[  \begin{array}{c}
\alpha \cr 
b_i \end{array} \Big]
\Big) R^{\alpha}NS^{F+\alpha}.
\end{equation}
Now we see that taking $\alpha =b_1$ (see eq. (\ref{eq18})) and 
$b_i=b_2, b_3$ for Susy/NonSusy models we get 
\begin{equation}
{\mathrm{Tr}}\, \frac{1}{2}\Big( 1\pm (-)^{b_3}
\Big) \times  \frac{1}{2} \Big( 1 - (-)^{b_2} \,
C\Big[  \begin{array}{c}
b_1 \cr b_2 \end{array} \Big]\, \Big)
R^{b_1}NS^{F+b_1}.
\end{equation}
This means that a massless state could be obtained by taking the
vacuum in the left sector (Ramond for the fermions in $b_1$
thus including the spacetime left fermions and NeveuSchwarz for
the other left fermions: thus getting a spacetime spinor)
and exciting one  NeveuSchwarz lowest mode of the 
spacetime right fermions (thus a spacetime vector). Because
$b_2$ contains the spacetime right fermions, all gravitino 
states are not projected out by the first projection if 
$C\Big[  \begin{array}{c} b_1 \cr  b_2 \end{array} \Big] =1$. On the other,
since $b_3$ does not contain spacetime right fermions this gravitino
state is projected out in the NonSusy model.

\section*{Appendix B: The classification of the two-loop terms 
into modular invariant sets}

Here we organize the various contributions in modular invariant sets,
that is we put in the same set contributions related by modular
transformations. We denote a one-loop spin structure by
\begin{equation}
\Big[  \begin{array}{c} a \cr b \end{array} \Big].
\end{equation}
A two-loops spin structure can be denoted by the tensor product of
two one-loop spin structures:
\begin{equation}
\Big[  \begin{array}{cc} a_1 & a_2 \cr 
b_1 & b_2 \end{array} \Big]
\sim
\Big[  \begin{array}{c} a_1 \cr 
b_1 \end{array} \Big] \times
\Big[  \begin{array}{c} a_2 \cr 
b_2 \end{array} \Big].
\end{equation} 
 
As we have seen, we have to consider the cases when 
$a (b)=\alpha + n b_{3,4} \,(\beta +\tilde n b_{3,4})$ where
$\alpha ,\beta \in \Xi'_0  = \{\emptyset, F + b_1 + b_2\}$. 
Let us for short take $\alpha =\beta =\emptyset$.
At one-loop the following three are related by modular transformations:
\begin{equation}
\Big[  \begin{array}{c} v \cr 
\emptyset \end{array} \Big],
\qquad  
\Big[  \begin{array}{c} \emptyset \cr 
v \end{array} \Big], 
\qquad 
\Big[  \begin{array}{c} v \cr 
v + F\end{array} \Big] \sim 
\Big[  \begin{array}{c} v \cr 
v \end{array} \Big],
\end{equation}
(we are interested in the case when $v=b_{3},b_{4},b_3+b_4$).
Thus we consider one representative say
$\Big[  \begin{array}{c} v \cr 
\emptyset \end{array} \Big]$
for the 3 elements. Similarly the various
$\Big[  \begin{array}{c} v \cr 
u \end{array} \Big]$
are all modular related, for $u \not= v$ and 
$u,v=b_{3},b_{4},b_3+b_4$, for a total of 6 elements:
we take as representative
$\Big[  \begin{array}{c} b_3 \cr 
b_4 \end{array} \Big]$.

The two-loops modular transformations (see below) give the relations,
for $u \not= v$ and $u,v=b_{3},b_{4}$:
\begin{eqnarray}
& & \Big[  \begin{array}{c} v \cr 
\emptyset \end{array} \Big] \times
\Big[  \begin{array}{c} v \cr 
\emptyset \end{array} \Big]
\sim
\Big[  \begin{array}{c} \emptyset \cr 
\emptyset \end{array} \Big] \times
\Big[  \begin{array}{c} v \cr 
\emptyset \end{array} \Big],
\\
& & \Big[  \begin{array}{c} v \cr 
\emptyset \end{array} \Big] \times
\Big[  \begin{array}{c} v+u \cr 
\emptyset \end{array} \Big]
\sim
\Big[  \begin{array}{c} v \cr 
\emptyset \end{array} \Big] \times
\Big[  \begin{array}{c} u \cr 
\emptyset \end{array} \Big], 
\\
& & \Big[  \begin{array}{c} v \cr 
\emptyset \end{array} \Big] \times
\Big[  \begin{array}{c} v \cr 
u \end{array} \Big]
\sim
\Big[  \begin{array}{c} \emptyset \cr 
\emptyset \end{array} \Big] \times
\Big[  \begin{array}{c} v \cr 
u \end{array} \Big], 
\\
& & \Big[  \begin{array}{c} v+u \cr 
\emptyset \end{array} \Big] \times
\Big[  \begin{array}{c} v \cr 
u \end{array} \Big]
\sim
\Big[  \begin{array}{c} \emptyset \cr 
\emptyset \end{array} \Big] \times
\Big[  \begin{array}{c} v \cr 
u \end{array} \Big],
\\ 
& & 
\Big[  \begin{array}{c} v \cr
u \end{array} \Big] \times
\Big[  \begin{array}{c} v \cr  
u \end{array} \Big]
\sim
\Big[  \begin{array}{c} v \cr
\emptyset \end{array} \Big] \times
\Big[  \begin{array}{c} u \cr
\emptyset \end{array} \Big].
\end{eqnarray}

In conclusion we get the 6 categories written in Sect.4--{\bf Organization of the
computation}, each category being a set of elements with the given 
multiplicity. For instance, 
$\Big[  \begin{array}{cc}\alpha_1 +b_3 & \alpha_2 \cr 
\beta_1 & \beta_2 \end{array} \Big]$ 
is obtained from
\begin{equation}
\Big[  \begin{array}{c} v \cr 
\emptyset \end{array} \Big] \times
\Big[  \begin{array}{c} v \cr 
\emptyset \end{array} \Big],  \qquad 
\Big[  \begin{array}{c} \emptyset \cr 
\emptyset \end{array} \Big] \times
\Big[  \begin{array}{c} v \cr 
\emptyset \end{array} \Big], \qquad 
\Big[  \begin{array}{c} v \cr 
\emptyset \end{array} \Big] \times
\Big[  \begin{array}{c} \emptyset \cr \emptyset \end{array} \Big]
\end{equation}
for a total of $3 \times 3+2\times 3=15$ modular related terms. For the other categories 
we have a similar counting: $2\times (6+3\times (3\times 6))=120$
terms for the category 5) and $6\times 6+2\times (3\times (3\times 3))=90$
terms for the category 6).

Finally, the two-loop modular transformation we have used above is:
\begin{equation}
Q \Big[  \begin{array}{cc} a_1 & a_2 \cr
b_1 & b_2    \end{array} \Big] \stackrel{\mathrm{X
}}{\longrightarrow}   \pm
Q \Big[  \begin{array}{cc} a_1 & a_2 \cr 
a_2+b_1  & a_1 + b_2   \end{array} \Big],
\end{equation}
which  would give the following modular transformation
 for  $V[\,]$:
\begin{equation}
V\Big[  \begin{array}{cc} \alpha_1 & \alpha_2 \cr
\beta_1 & \beta_2    \end{array} \Big]  
\stackrel{\mathrm{X}}{\longrightarrow}
V \Big[  \begin{array}{cc} \alpha_1 & \alpha_2  \cr
\alpha_2 + \beta_1 & \alpha_1 + \beta_2  
    \end{array} \Big].
\end{equation}
The other generating modular transformations are
 just one-loop like modular
transformations which change only one column
of the characteristics of the  theta function, i.e. $Q_{s}$.
These are denoted as $T_{1,2}$ and $S_{1,2}$ in an obvious notation.   
The following example demonstrates how we change
$V\Big[  \begin{array}{cc}
\alpha_1  + b_3 & \alpha_2 + b_4 \cr
\beta_1 & \beta_2 \end{array} \Big] $ to
$V\Big[  \begin{array}{cc} \tilde{\alpha}_1  + b_4 &
\tilde{\alpha}_2 + b_3  \cr   
\tilde{\beta}_1 & \tilde{\beta}_2
\end{array} \Big] $ by a series of modular
transformations:
\begin{eqnarray}
V\Big[  \begin{array}{cc}
\alpha_1  + b_3 & \alpha_2 + b_4 \cr
\beta_1 & \beta_2 \end{array} \Big] 
& \stackrel{\mathrm{X}}{\longrightarrow} &
V\Big[  \begin{array}{cc} \alpha_1
+ b_3 & \alpha_2 + b_4 \cr    
\alpha_2  + \beta_1 + b_4 & \alpha_1 +
 \beta_2  + b_3  \end{array} \Big]
\nonumber \\
& \stackrel{\mathrm{S_1S_2}}{\longrightarrow} &
V\Big[  \begin{array}{cc} \alpha_2  + \beta_1
+ b_4 & \alpha_1 + \beta_2  + b_3  \cr
\alpha_1  + b_3 & \alpha_2 + b_4 \end{array} \Big]
\nonumber \\
& \stackrel{\mathrm{X}}{\longrightarrow} &
V\Big[  \begin{array}{cc}  \alpha_2  + \beta_1
 + b_4 & \alpha_1 + \beta_2  + b_3
  \cr\beta_2 & \beta_1
\end{array} \Big].
\end{eqnarray}

\section*{Appendix C: Two-loop vanishing of the cosmological
constant in supersymmetric string models}

First let us show that  $V\Big[  
\begin{array}{cc} \alpha_1 & \alpha_2\cr
\beta_1 & \beta_2   \end{array} \Big]  =0$. 
 Because $\alpha_i, \beta_j \in \Xi'_0, i,
j =1,2 $, we have  $b_k\cap \alpha_i = b_k 
\cap \beta_j = \emptyset, k=1,2$, and  
\begin{eqnarray}
 & & V 
\Big[  \begin{array}{cc} \alpha_1 & \alpha_2\cr
\beta_1 & \beta_2   \end{array} \Big]    \propto   
\sum_{n_i,m_i,\tilde{n}_i, \tilde{m}_i=0}^1 
C\Big[  \begin{array}{cc} \alpha_1 
+n_1 b_1 + m_1 b_2 & \alpha_2
+n_2 b_1 + m_2 b_2 \cr \beta_1 
+\tilde{n}_1 b_1  + \tilde{m}_1 b_2 & \beta_2 
  +\tilde{n}_2 b_1  + \tilde{m}_2 b_2 \end{array} 
\Big]  
\nonumber \\
& & \quad \times \langle J(r+) J(r-) \tilde{J}(\bar{s}
+)\tilde{J}(\bar{s}+) 
\rangle_{s(\psi,\tilde{\psi})} \, Q \Big[ 
 \begin{array}{cc} n_1  & n_2 \cr
\tilde{n}_1  & \tilde{n}_2  \end{array} \Big]  
\, \bar{Q} \Big[  \begin{array}{cc} m_1  & m_2 \cr
\tilde{m}_1  & \tilde{m}_2  \end{array} \Big]  .
\label{sumaa}
\end{eqnarray}
Noticing the following result for the coefficients $C$:
\begin{eqnarray} 
& & C\Big[  \begin{array}{cc} \alpha_1 
+n_1 b_1 + m_1 b_2 & \alpha_2
+n_2 b_1 + m_2 b_2 \cr
\beta_1 
+\tilde{n}_1 b_1  + \tilde{m}_1 b_2 & \beta_2 
  +\tilde{n}_2 b_1  + \tilde{m}_2 b_2 \end{array} \Big]    
\nonumber \\
& & \qquad =  
C\Big[  \begin{array}{c  } \alpha_1 
+n_1 b_1 + m_1 b_2 \cr \beta_1 
+\tilde{n}_1 b_1  + \tilde{m}_1 b_2 \end{array} \Big]
C\Big[  \begin{array}{ c} \alpha_2 
+n_2 b_1 + m_2 b_2 \cr
\beta_2   +\tilde{n}_2 b_1  + \tilde{m}_2 b_2
 \end{array} \Big]   
\nonumber  \\
& & \qquad  = C\Big[  \begin{array}{c  } \alpha_1 
\cr \beta_1  \end{array} \Big]  
C\Big[  \begin{array}{c  } \alpha_2 
\cr \beta_2  \end{array} \Big]  \, (-1)^{n_1 + n_2 +
 \tilde{n}_1 + \tilde{n}_2 } 
\, (-1)^{m_1 + m_2 + \tilde{m}_1 + \tilde{m}_2 }  , 
\end{eqnarray}
the summation over the spin structure dependent parts 
in (\ref{sumaa}) is 
factorized into left and right parts. For the left part we have 
\begin{eqnarray}
V_L & & \hskip -1cm \Big[  \begin{array}{cc} \alpha_1 
& \alpha_2\cr
\beta_1 & \beta_2   \end{array} \Big]    
 \propto  
\sum_{n_i,\tilde{n}_i=0}^1 
\langle J(r+) J(r-)\rangle_{s(\psi,\chi^I,y^I,\omega^I)} \, 
(-1)^{n_1 +n_2 +\tilde{n}_1 +\tilde{n}_2} \, Q
\Big[  \begin{array}{cc} n_1 & n_2 \cr
\tilde{n}_1 &  \tilde{n}_2    \end{array} \Big]    
\nonumber \\
& =  &  \sum_{n_i,\tilde{n}_i=0}^1 
 (-1)^{n_1 +n_2 +\tilde{n}_1 +\tilde{n}_2} 
\left( K_1 Y_1 \Big[  \begin{array}{cc} n_1 & n_2 \cr
\tilde{n}_1 &  \tilde{n}_2    \end{array} \Big]     + 
K_2 Y_2 \Big[  
\begin{array}{cc} n_1 & n_2 \cr
\tilde{n}_1 &  \tilde{n}_2    \end{array} \Big]     \right. 
\nonumber \\
&  & \quad \left. - \sum_{I=1}^6 Y_1^2 
\Big[  \begin{array}{cc} \alpha_1(y^I) & \alpha_2 (y^I)  \cr
\beta_1(y^I) & \beta_2(y^I)    \end{array} \Big]    Y_1 \Big[ 
 \begin{array}{cc} n_1 & n_2 \cr
\tilde{n}_1 &  \tilde{n}_2    \end{array} \Big]     
\right) Q\Big[  \begin{array}{cc} n_1 & n_2 \cr
\tilde{n}_1 &  \tilde{n}_2    \end{array} \Big]     
\nonumber \\
& = & 0, 
\end{eqnarray}
which was proved in \cite{GavaIengoSotkov}. 

The second is to prove $V 
\Big[  \begin{array}{cc} \alpha_1 + b_3 & \alpha_2\cr
\beta_1 & \beta_2   \end{array} \Big]  =0$. We have 
\begin{eqnarray}
& &  V\Big[  \begin{array}{cc} 
\alpha_1 + b_3 & \alpha_2\cr
\beta_1 & \beta_2   \end{array} \Big]   \propto  
\sum_{n_i,m_i,\tilde{n}_i, \tilde{m}_i=0}^1   
\langle J(r+) J(r-) \tilde{J}(\bar{s}+)\tilde{J}(\bar{s}+) 
\rangle_{s(\psi,\tilde{\psi})}
\nonumber \\
& & \quad \times 
C\Big[  \begin{array}{cc} \alpha_1  + b_3 
+n_1 b_1 + m_1 b_2 & \alpha_2
+n_2 b_1 + m_2 b_2 \cr \beta_1 
+\tilde{n}_1 b_1  + \tilde{m}_1 b_2 & \beta_2 
  +\tilde{n}_2 b_1  + \tilde{m}_2 b_2 \end{array} \Big]
\nonumber \\
& & \quad \times 
Q^{1/2} \Big[  \begin{array}{cc} n_1  & n_2 \cr
\tilde{n}_1  & \tilde{n}_2  \end{array} \Big] 
Q^{1/2} \Big[  \begin{array}{cc} n_1 +1  & n_2 \cr
\tilde{n}_1  & \tilde{n}_2  \end{array} \Big] 
\, \bar{Q} \Big[  \begin{array}{cc} m_1  & m_2 \cr
\tilde{m}_1  & \tilde{m}_2  \end{array} \Big]  .
\label{sumbb}
\end{eqnarray}
Now we need the following coefficient:
\begin{eqnarray} 
& & C\Big[  \begin{array}{cc} \alpha_1  + b_3 
+n_1 b_1 + m_1 b_2 & \alpha_2
+n_2 b_1 + m_2 b_2 \cr \beta_1  
+\tilde{n}_1 b_1  + \tilde{m}_1 b_2 & \beta_2 
  +\tilde{n}_2 b_1  + \tilde{m}_2 b_2 \end{array} \Big]    
\nonumber \\
& & \qquad =  
C\Big[  \begin{array}{c  } \alpha_1 + b_3 
+n_1 b_1 + m_1 b_2 \cr \beta_1 
+\tilde{n}_1 b_1  + \tilde{m}_1 b_2 \end{array} \Big]
C\Big[  \begin{array}{ c} \alpha_2 
+n_2 b_1 + m_2 b_2 \cr
\beta_2   +\tilde{n}_2 b_1  + \tilde{m}_2 b_2 \end{array} \Big]   
\nonumber  \\
& & \qquad 
 = (-1)^{ n_2 + \tilde{n}_2 + m_2 + \tilde{m}_2 }  \,  C\Big[ 
 \begin{array}{c  } \alpha_1  + b_3 
\cr \beta_1  \end{array} \Big]  
C\Big[  \begin{array}{c  } \alpha_2 
\cr \beta_2  \end{array} \Big]  
\nonumber \\
& & \qquad \qquad \times 
C\Big[  \begin{array}{c  }  b_3  + n_1 b_1 
\cr \tilde{n}_1 b_1   \end{array} \Big]  \,
C\Big[  \begin{array}{c  }  b_3  + m_1 b_2 
\cr \tilde{m}_1 b_2   \end{array} \Big]  .
\end{eqnarray}
Again we find a splitting of the left and right parts in 
eq. (\ref{sumbb}). 
For the left part we have 
\begin{eqnarray}
& & \hskip -.5cm V_L  \Big[  \begin{array}{cc} \alpha_1  
+ b_3 & \alpha_2\cr
\beta_1 & \beta_2   \end{array} \Big]    
 \propto  
\sum_{n_i,\tilde{n}_i=0}^1 
(-1)^{n_2 +\tilde{n}_2} \, 
\langle J(r+) J(r-)\rangle_{s(\psi,\chi^I,y^I,\omega^I)} 
\nonumber \\
& & \quad \times 
C\Big[  \begin{array}{c  }  b_3  + n_1 b_1 
\cr \tilde{n}_1 b_1   \end{array} \Big]  \, Q^{1/2}
\Big[  \begin{array}{cc} n_1 & n_2 \cr
\tilde{n}_1 &  \tilde{n}_2    \end{array} \Big]    
\, Q^{1/2} \Big[  \begin{array}{cc} n_1 + 1  & n_2 \cr
\tilde{n}_1 &  \tilde{n}_2    \end{array} \Big]  .
\end{eqnarray}
In the above there appears two different spin structures in $Q$. 
These two spin structures should both be even in order to have a
 possibly non-vanishing  result. This requires $\tilde{n}_1 = 0$ 
and we have 
\begin{eqnarray}
V_L & & \hskip -1cm \Big[  \begin{array}{cc} \alpha_1  +
 b_3 & \alpha_2\cr
\beta_1 & \beta_2   \end{array} \Big]    
 \propto  
\sum_{n_1, n_2, \tilde{n}_2 =0}^1 
(-1)^{n_1 + n_2 +\tilde{n}_2} \, 
\langle J(r+) J(r-)\rangle_{s(\psi,\chi^I,y^I,\omega^I)} 
\nonumber \\
& & \quad \times 
Q^{1/2}  \Big[  \begin{array}{cc} n_1 & n_2 \cr
0  &  \tilde{n}_2    \end{array} \Big]    
\, Q^{1/2} \Big[  \begin{array}{cc} n_1 + 1  & n_2 \cr
0 &  \tilde{n}_2    \end{array} \Big] 
\nonumber  \\
& =  &  \sum_{n_1, n_2, \tilde{n}_2=0}^1 
 (-1)^{n_1 +n_2+\tilde{n}_2} 
\left( K_1 Y_1 \Big[  \begin{array}{cc} n_1 & n_2 \cr
0 &  \tilde{n}_2    \end{array} \Big]     + K_2 Y_2 \Big[  
\begin{array}{cc} n_1 & n_2 \cr
0 &  \tilde{n}_2    \end{array} \Big]     \right. 
\nonumber \\
&  & \quad  - \sum_{I=1}^6 Y_1 
\Big[  \begin{array}{cc} (\alpha_1 + b_3) 
(y^I) & \alpha_2 (y^I)  \cr
\beta_1(y^I) & \beta_2(y^I)    \end{array} \Big]    
Y_1  \Big[  \begin{array}{cc} (\alpha_1 + b_3) 
(\omega^I) & \alpha_2 (\omega^I)  \cr
\beta_1(\omega^I) & \beta_2(\omega^I)    \end{array} \Big]    
\nonumber \\
& & \quad \times \left. 
Y_1 \Big[  \begin{array}{cc} n_1 & n_2 \cr
0 &  \tilde{n}_2    \end{array} \Big]     
\right) 
Q^{1/2}  \Big[  \begin{array}{cc} n_1 & n_2 \cr
0  &  \tilde{n}_2    \end{array} \Big]    
\, Q^{1/2} \Big[  \begin{array}{cc} n_1 + 1  & n_2 \cr
0 &  \tilde{n}_2    \end{array} \Big] 
\nonumber \\
&  = & 0, 
\end{eqnarray}
by making use of the following identities ($i=1,2$):
\begin{eqnarray}
 & & \left( Y_i \Big[  \begin{array}{cc} 0 & 0 \cr
0  &  0    \end{array} \Big]  - 
Y_i    \Big[  \begin{array}{cc} 1 & 0 \cr
0  &  0    \end{array} \Big]  \right) 
Q^{1/2} \Big[  \begin{array}{cc} 0 & 0 \cr
0  &  0    \end{array} \Big]  \, 
Q^{1/2} \Big[  \begin{array}{cc} 1 & 0 \cr
0  &  0    \end{array} \Big]  
\nonumber \\
& & \quad 
- \left( Y_i \Big[  \begin{array}{cc} 0 & 1 \cr
0  &  0    \end{array} \Big]  - 
Y_i    \Big[  \begin{array}{cc} 1 & 1 \cr
0  &  0    \end{array} \Big]  \right) 
Q^{1/2} \Big[  \begin{array}{cc} 0 & 1 \cr
0  &  0    \end{array} \Big]  \, 
Q^{1/2} \Big[  \begin{array}{cc} 1 & 1\cr
0  &  0    \end{array} \Big]  
\nonumber \\
& & \quad -\left( Y_i \Big[  \begin{array}{cc} 0 & 0 \cr
0  &  1    \end{array} \Big]  - 
Y_i    \Big[  \begin{array}{cc} 1 & 0 \cr
0  &  1    \end{array} \Big]  \right) 
Q^{1/2} \Big[  \begin{array}{cc} 0 & 0 \cr
0  &  1    \end{array} \Big]  \, 
Q^{1/2} \Big[  \begin{array}{cc} 1 & 0 \cr
0  &  1    \end{array} \Big]   = 0, 
\label{diffaaa}
\end{eqnarray}
which was proved explicitly in \cite{IengoGalenI}. 
Similar method can be used to prove the vanishing of all 
$V[\,]$'s in the 3rd and 4th categories.

\section *{Acknowledgments}

We would like to tkank E. Gava, G. Ferretti and
K.S. Narain for helpful discussions. 
C.-J. Zhu is supported in part by funds from National 
Natural Science  Foundation of China and Pandeng Project.
He would also like 
to thank the hospitality of SISSA where most of the work 
was done.

\end{document}